\documentclass[12pt]{modifiediopart}

\usepackage{graphicx}

\renewcommand{\theequation}{\thesection.\arabic{equation}}

\newcommand{\CS}{c_{\mathrm s}}
\newcommand{\CSO}{c_{{\mathrm s}0}}
\newcommand{\EMAX}{\varepsilon_{\mathrm {max}}}

\begin{document} 

\title{Probing the stability of gravastars by dropping dust shells onto them}

\author{Merse El\H{o}d G\'asp\'ar and  Istv\'an R\'acz}

\address{\small RMKI, 
\small  H-1121 Budapest, Konkoly Thege Mikl\'os \'ut 29-33.
\small Hungary}

\ead{merse@rmki.kfki.hu, iracz@rmki.kfki.hu}

\begin{abstract}
As a preparation for the dynamical investigations, 
this paper begins with a short review of the three-layer gravastar model 
with distinguished attention to the structure of the pertinent parameter space of gravastars in equilibrium.
Then the radial stability of these types of gravastars is studied 
by determining their response for the totally inelastic collision of their surface layer with a dust shell. 
It is assumed that the dominant energy condition holds 
and the speed of sound does not exceed that of the light in the matter of the surface layer. 
While in the analytic setup the equation of state is kept to be generic, 
in the numerical investigations three functionally distinct classes of equations of states are applied. 
In the corresponding particular cases the maximal mass of the dust shell 
that may fall onto a gravastar without converting it into a black hole is determined. 
For those configurations which remain stable the excursion of their radius is assigned. 
It is found that even the most compact gravastars cannot get beyond the lower limit of the size of conventional stars, 
provided that the dominant energy condition holds in both cases. 
It is also shown---independent of any assumption concerning the matter 
interbridging the internal de Sitter and the external Schwarzschild regions---that the better is 
a gravastar in mimicking a black hole the easier is to get the system formed by a dust shell 
and the gravastar beyond the event horizon of the composite system. 
In addition, a generic description of the totally inelastic collision of spherical shells 
in spherically symmetric spacetimes is also provided in the appendix.
\end{abstract}


\section{Introduction}

There are more and more astrophysical observations justifying the existence of extremely compact massive objects 
with size close to their Schwarzschild radius \cite{Celotti99,Kormendy03,Melia05}. 
It is widely accepted that these observations also provide indirect justifications of the existence of black holes (BHs). 
Nevertheless, there are also alternative ideas about claiming that exotic states of matter may exist 
which could stabilize extremely compact stars suiting to the aforementioned astrophysical observations (see, e.g., \cite{Visser09} for a recent review).  
One of the most popular among these types of BH mimicking objects is the gravitational vacuum star (gravastar) model 
which has received considerable attention not least because its relation to the concept of dark energy. 
In this model of Mazur and Mottola \cite{Mazur01} an interior de Sitter spacetime region is connected via three intermediate layers 
to an outer Schwarzschild solution such that the radius of the outermost layer is supposed to be slightly larger than the Schwarzschild radius of the system. 

It is worth mentioning that in advance to the gravastar model there were several constructions 
in which the matching of a de Sitter region to the Schwarzschild spacetime was applied. 
For instance, to get rid of the $r=0$ singularity of the Schwarzschild spacetime, Frolov, Markov and Mukhanov in \cite{Frolov88} 
proposed a matching of a de Sitter interior to it at a small radius of the Planck scale 
ensuring thereby that the curvature remains bounded everywhere in the yielded spacetime. 
Dynamical investigation of this model was already carried out in \cite{Balbinot90}. 

Note that up to certain extent the model used in \cite{Frolov88} could be considered 
as the precursor of the gravastar model---although in the latter the matching was made in a  more elaborated way---and the outermost matching surface, 
at the boundary of the Schwarzschild region, was supposed to be arranged 
such that its radius is slightly larger than the pertinent Schwarzschild radius.

Once such a model is set up the following questions manifest themselves: 
\begin{itemize}
\item[(1)] What type of physical process may produce such a gravastar?
\item[(2)] Is a gravastar stable?
\item[(3)] If it is, does it provide a viable alternative to BHs?
\end{itemize}

While the first question has not been tackled yet even the second question turned out to be too complex 
within the original model of Mazur and Mottola---although in \cite{Mazur01} an argument claiming for 
its thermodynamical stability was given---as it is composed by making use of three different types of regions with unspecified matter. 
To reduce the related ambiguities Visser and Wiltshire \cite{Visser04} introduced a simplified 
three-layer gravastar model where the interior de Sitter region is matched to the exterior Schwarzschild spacetime via a single matter shell. 
This model is simple enough to carry out various analytic investigations by making use of the thin-shell formalism of Israel \cite{Israel66}. 
Visser and Wiltshire besides deriving the basic relations determining the evolution of gravastars also carried out the first investigation of their radial stability.  
Since then the stability of gravastars has been studied by several authors within this simplified model or within its continuum correspondence \cite{Cattoen05}. 
Results relevant for radial stability can be found in \cite{Carter05,Bilic00}, and in case of electrically charged gravastars in \cite{Horvat09}. 
The stability has also been investigated with respect to axial perturbations \cite{DeBenedictis06,Chirenti07}.

In all of these investigations attention was restricted to the space of gravastars in equilibrium, 
i.e. the radial stability was investigated by determining the response of a gravastar to a slight formal change of the underlying effective potential. 
For instance, in \cite{Rocha08a, Rocha08b} the excursions of gravastars was investigated in such a way that their evolution started with carefully prepared initial conditions.
In all of the pertinent investigations it was demonstrated that by suitably adjusting the equation of state (EOS) of the matter 
forming the surface of the gravastar the subspace of `stable' gravastars may always be ensured to be of non-zero measure. 
Nevertheless, it was also found that whenever the measure of the subspaces of the configuration space 
representing stable and unstable gravastars is compared the former is always found to be negligible with respect to the latter. 
This observation was commonly interpreted that gravastars may not offer a viable alternative to BHs.

The main purpose of the present paper is to determine the response of a gravastar in equilibrium to the arrival of a dust shell onto its surface. 
This is done not merely by considering some formal change of the effective potential determining the state of a gravastar but also by making use of the full dynamical setup. 
For the sake of definiteness, we assume that the surface of the gravastar and the dust shell collide in a totally inelastic manner. 
In addition, concrete EOSs are chosen and it is assumed that the dominant energy condition (DEC) holds\footnote{Recall that 
the dominant energy condition guarantees that the concept of causality is properly adopted in general relativity. 
Therefore, the exclusion of matter models yielding its violation seems to be preferable.} and the speed of sound in the surface of the gravastar 
does not exceed that of the light. Then the relevant non-linear problems---the basic equations 
of which are based on the dynamics of spherical shells---are solved by using analytical and numerical approaches. 
In this way we can not only study the excursion of particular gravastar models 
but also determine the maximal mass of the dust shell colliding with the surface of the gravastar without converting the latter into a BH.

This paper is organized as follows. 
In section \ref{Visser-model} some of the basics of the Visser and Wiltshire three-layer dynamics gravastar model
are recalled using dimensionless variables. 
As a preparation for the aforementioned dynamical investigations a short survey of the configuration space of stable gravastars is also provided. 
(Although there are no completely new results in this section, 
we believe that this review provides a good reference frame for the results of the succeeding dynamical investigations.) 
In Section \ref{Collision-section} the dynamics of the system composed by the spherically symmetric dust shell 
falling onto a stable gravastar, along with their collision, is described. 
Section \ref{Results} reports about our analytical and numerical results concerning the dynamics of maximally loaded gravastars, 
while Section \ref{Conclusions} contains our concluding remarks. Finally, in the appendix, 
a generic description of the totally inelastic collision of spherical shells in spherically symmetric spacetimes is provided. 
Throughout this paper the geometrized units, with $G = c= 1$, are applied.

\section{Gravastar model of Visser and Wiltshire}
\label{Visser-model}

Throughout this paper considerations will be restricted to the three-layer
spherically symmetric gravastar model of Visser and Wiltshire
\cite{Visser04}. This simplified model consists of an external Schwarzschild
vacuum region with the mass parameter $M$, representing the total ADM mass of the
gravastar, an interior de Sitter region with energy density $\rho_0$, and
an infinitesimally thin shell at the mutual boundary of the aforementioned two
regions at radius $a$  with surface energy density $\sigma$ and surface
tension $\theta$. The line element representing the interior de\,Sitter 
and exterior Schwarzschild spacetimes is given in the form of (\ref{met1}) and (\ref{met2}), 
where the pertinent forms of $M_K(r)$ are specified in the paragraph below (\ref{met2}).
Note that the radius is allowed to vary in time. As the 
shell is assumed to coincide, at any moment, with one
of the $SO(3)$ group orbits, the quantities characterizing the shell are only the
functions of the radius. To have a closed system of equations governing
the evolution of the shell we also need to assume the existence of an EOS. 
As was noted above $\sigma$
and $\theta$ must be functions of the radius exclusively. Thereby, in virtue
of the implicit function theorem, the EOS, relating the tension, $\theta$, of
the surface to the mass density, $\sigma$, of the surface of the gravastar,
may always be assumed to possess the form $\theta=\theta(\sigma)$.  This
section, besides recalling the most important equations determining the
dynamics, also provides a short comprehensive review of the basic
properties of the above-outlined three-layer gravastar model. 

\subsection{Parametrization}

In studying the gravastar model the use of the following dimensionless
parameters turned out to be advantageous. They are the dimensionless radius
$\alpha = a/(2M)$,  surface density $\Sigma = \sigma M$, tension $\Theta =
\theta M$ and the dimensionless de Sitter density parameter, $\eta = 8 k M^2$,
where $k=4\pi\rho_0/3$.\footnote{Whenever $\eta =k=0$ 
the systems degenerates to a shell moving in a Minkowski-Schwarzschild spacetime.
\vspace*{2.5mm}} 
Note that in the chosen setup positive tangential pressure does
correspond to negative surface tension. To avoid the appearance of the
event horizon  in the Schwarzschild region the inequality $1 < \alpha$ is
required to be satisfied. Similarly, in order to exclude the presence of a
`cosmological' horizon in the de Sitter region the inequality $\alpha <
\eta^{-1/2}$ is imposed. Accordingly, in case of the considered gravastar
models the value of $\eta$ is restricted to the interval $0 < \eta < 1$. 
Finally note that in dynamical situations the time dependence may conveniently be
expressed via the dimensionless proper time measured by observers moving
radially along the world-sheet of the shell, in $2M$ units.

\subsection{Dynamical equations}

The basic equations determining the dynamics of the shell of the gravastar can
be derived in various  ways.  The most widely applied method 
uses the thin-shell formalism of Israel \cite{Israel66} based on the results given by Sen, Lanczos, Darmois
\cite{Sen24,Lanczos24,Darmois27}. In this approach the induced metrics, $h_{AB}^-$
and $h_{AB}^+$, on the mutual boundary of the two spacetime regions---the inner
and the outer---on the two sides of the shell are assumed to coincide, 
i.e. the metric on the shell can be given as $h_{AB}=h_{AB}^+=h_{AB}^-$, 
while the discontinuity of the pertinent extrinsic curvature tensors, $K_{AB}^+$ and $K_{AB}^-$, 
are related to the surface--energy density $S_{AB} = \mathrm{diag}(\sigma,-\theta,-\theta)$ via
the matching condition\footnote{This relation corresponds essentially to the Lanczos equation.} 

\begin{equation}
\label{matchcond}
K_{AB}^+-K_{AB}^-=-8\pi\left\{S_{AB}-h_{AB}
(h^{CD}S_{CD})\right\} \, .
\end{equation} 

By making use of the pertinent results of \cite{Visser04}, along with the
above-introduced parameterization, these junction conditions can be seen to
take the form 

\vspace*{-4mm}
\begin{eqnarray}
\label{sigma-dyn}
\Sigma &=&
\frac{1}{8\pi\alpha}\left(\sqrt{1-\eta\alpha^2+\dot{\alpha}^2}
-\sqrt{1-\alpha^{-1}+\dot{\alpha}^2}\;\right) \, , \\
\label{theta-dyn}
\Theta &=&
\frac{1}{16\pi\alpha}\left(\frac{1-2\eta\alpha^2+\dot{\alpha}^2
  +\alpha\ddot{\alpha}}{\sqrt{1-\eta\alpha^2+\dot{\alpha}^2}} 
-\frac{1-\alpha^{-1}/2+\dot{\alpha}^2 +\alpha\ddot{\alpha}}{\sqrt{1-\alpha^{-1}
    +\dot{\alpha}^2}}\right) \, .
\end{eqnarray}

\noindent
where the over-dot denotes the derivative with respect to the dimensionless
proper time along the shell. 

As the energy density of the shell is always assumed to be non-negative, the
first square root on the left-hand side of (\ref{sigma-dyn})  must be greater
than the second one, which implies that $\alpha < \eta^{-1/3}$. Note that this
restriction, viewed as a condition on the value of $\alpha$, is more demanding
than the previous one excluding the presence of a cosmological horizon in the
inner de Sitter region.  In virtue of (\ref{sigma-dyn}) it is straightforward
to see that whenever $\Sigma$ is non-negative it attains its maximum
value  for vanishing $\dot{\alpha}$ for any fixed value of $\alpha$. 

In describing the dynamics of gravastars it turned out to be advantageous to
introduce an `effective potential' defined via the relation 

\begin{equation}
\label{eqV}
V(\alpha; \eta)=-\frac{\dot{\alpha}^2}{2} \; .
\end{equation}

\noindent
Assuming that $\Sigma$ is positive, by a straightforward
algebraic manipulation, one finds that $V$ can be given as a function of
$\alpha$ and $\Sigma=\Sigma(\alpha)$---it also depends on the dimensionless de
Sitter density parameter $\eta$---and it reads

\begin{equation}
\label{V}
\hskip-0.5cm V(\alpha,\Sigma(\alpha); \eta) = \frac{1}{2}\left[1+
  \frac{\eta}{64\pi^2\alpha\,\Sigma^2(\alpha)}- 
  \left(4\pi\alpha\,\Sigma(\alpha)+
  \frac{1+\eta\alpha^3}{16\pi\alpha^2\,\Sigma(\alpha)}\right)^2 
  \right] \, ,
\end{equation}

\noindent
which, in virtue of (\ref{eqV})---despite its complicated analytic form---has to be 
smaller than or equal to zero. 

Once $\Sigma$ is known as a function of the dimensionless radius, $\alpha$, so
is $V(\alpha; \eta)$, and, in turn, (\ref{eqV}) can be used to determine the
motion of the shell. The next subsection shows that the EOS of the shell,
along with the conservation law, $D^AS_{AB}=0$---where $D_A$ denotes the
covariant derivative operator compatible with the metric, $h_{AB}$, on the
shell---can be used to determine the functional form of $\Sigma(\alpha)$. 

\subsection{Equation of state of the shell}
\label{EOSsubsec}

Recall first that the EOS of the surface matter of the gravastar, if given in terms 
of the dimensionless variables, may be assumed to possess the form
$\Theta=\Theta(\Sigma)$.  Admittedly, the entire concept of an infinitesimally
thin shell may be considered to be somewhat artificial. 
Nevertheless, in our investigations we need to apply some assumptions on the type of its matter.
The use of a suitable energy condition is essential. In this respect the use of the
DEC seems  to be the most appropriate which is supposed to guarantee that the concept of causality is properly adopted in
describing the evolution of coupled gravity matter systems.  Note that the
less stringent weak and null energy conditions are automatically satisfied
whenever the DEC is guaranteed to hold. For an infinitesimally thin shell the DEC can
be seen to hold  whenever $|\Theta|<\Sigma$ \cite{barcelo02}.

Throughout this paper the function $\Theta=\Theta(\Sigma)$ will be assumed to
be continuous and piecewise differentiable. To avoid dynamical instabilities we shall also
assume that the square of the speed of sound, $\CS^2=-\rmd\Theta/\rmd\Sigma$,
is non-negative, and to be compatible with the concept of relativity that $\CS^2$ is 
required to be less than or equal to the square of the
speed of light.  Accordingly, unless otherwise stated, it will be assumed that
$0\leq \CS^2\leq 1$.

It is well-known that the radial conservation equation is

\begin{equation}
\label{conserv}
\frac{\rmd}{\rmd\tau}\left(\Sigma \alpha^2\right) =
\Theta\frac{\rmd}{\rmd\tau}\left(\alpha^2\right) ,
\end{equation}

\noindent
where $\tau$ denotes the dimensionless proper time along the radial trajectories generating
the   shell. (\ref{conserv}) may be derived either by referring the
conservation law $D^AS_{AB}=0$ or by making use of equations (\ref{sigma-dyn})
and (\ref{theta-dyn}) as in \cite{Visser04}. By introducing the dimensionless
radius $\alpha$, instead of $\tau$, as an independent variable (\ref{conserv})
takes the form 

\begin{equation}
\label{dsigma}
\alpha\frac{\rmd\Sigma}{\rmd \alpha} = 2(\Theta-\Sigma) \, .
\end{equation}

\noindent
This is the very point where the EOS of the shell comes into play since by
substituting the relation $\Theta=\Theta(\Sigma)$ into (\ref{dsigma}) the
functional form of $\Sigma=\Sigma(\alpha)$ can be determined. Note that as
$\Theta-\Sigma<0$, in virtue of (\ref{dsigma}), $\Sigma(\alpha)$ has to be a
monotonic decreasing function.
Since (\ref{dsigma}) is a separable differential equation,
an implicit solution of it can be given as 

\begin{equation}\label{general-eos}
\frac{\alpha}{\alpha_0} = \exp{\left(\frac{1}{2}\int_{\Sigma_0}^{\Sigma(\alpha)}
\frac{\rmd \widetilde{\Sigma}}{\Theta(\widetilde{\Sigma}) - \widetilde{\Sigma}}\right)} \, ,
\end{equation}

\noindent
where the integration constant $\Sigma_0$ is fixed such that
$\Sigma_0=\Sigma(\alpha_0)$.\footnote{Throughout this subsection the integration constant are determined by referring to certain $\alpha_0$ value. 
Although latter $\alpha_0$ will label equilibrium states, 
it should keep in mind that $\alpha_0$ does not have such an interpretation yet.}

To have some specific examples in later sections we shall use the following three types of EOS. 
The simplest EOS---that has been used in most of the former investigations of gravastars---is linear 
homogeneous possessing the functional form $\Theta(\Sigma)=-w_0\,\Sigma$, 
which, however, is not satisfied for a generic linear functional relation. In this case $\CS^2=w_0$ and

\begin{equation}
\label{homogeneousEOS}
\Sigma(\alpha) = \Sigma_0\,\left(\frac{\alpha}{\alpha_0}\right)^{-2(1+w_0)} .
\end{equation}

Clearly, it is of interest to consider more complex systems, as well. 
Since ${\rmd\Theta}/{\rmd\Sigma}$ is related to the speed of sound and its value at an equilibrium point will play an important role 
in the stability of the gravastar, it seems to be reasonable consider the EOS with varying sound speed at the equilibrium point. 
The simplest generalization of a homogeneous linear EOS
would be the non-homogeneous linear one. However, we would also like to have the DEC to hold which requires $\Theta$ to vanish at $\Sigma=0$. 
Therefore we use instead the `broken linear' EOS 

\begin{equation}
\label{truncatedEOS}
\Theta(\Sigma) = 
\left\{
\begin{array}{ll}
-w_1\Sigma & \mbox{, if $\Sigma \leq \Sigma_1$} \\
-w_1\Sigma_1-w_2(\Sigma-\Sigma_1) & \mbox{, if $\Sigma > \Sigma_1$} \; .
\end{array}
\right.
\end{equation}

\noindent
Here the parameter $\Sigma_1$ signifies the value of the breaking point in the $\Sigma$ range, 
while for the values of the slopes, $w_1$ and $w_2$, 
before and after the matching the inequalities $0\leq w_1, w_2 < 1$ hold.
Then (\ref{general-eos}) can be evaluated explicitly. 
In doing so the dimensionless radius of the matching point $\alpha_1$, 
gets to be determined by  the relation $\Sigma_1=\Sigma(\alpha_1)$ as

\begin{equation}
\label{alpha1}
\alpha_1 = 
\left\{
\begin{array}{ll}
\alpha_0 \left( \frac{\Sigma_0}{\Sigma_1}\right)^{1/(2w_1+2)} & \mbox{, if $\Sigma_0 \leq \Sigma_1$} \\
\alpha_0 \left( \frac{\Sigma_0(1+w_2)+\Sigma_1(w_1-w_2)}{\Sigma_1(1+w_1)}\right)^{1/(2w_2+2)} & \mbox{, if $\Sigma_0 > \Sigma_1$} \; .
\end{array}
\right.
\end{equation}

\noindent
Then, whenever $\Sigma_0 \leq \Sigma_1$

\begin{equation}
\Sigma(\alpha)= 
\left\{
\begin{array}{ll}
\Sigma_1\frac{w_2-w_1+(1+w_1)\left( \frac{\alpha_1}{\alpha}\right)^{2w_2+2} }{1+w_2} & \mbox{, if $\alpha \leq \alpha_1$} \\
\Sigma_0\left(\frac{\alpha_0}{\alpha}\right)^{2w_1+2\phantom{{}^{^\frac{}{}}}} & \mbox{, if $\alpha > \alpha_1$} \; ,
\end{array}
\right.
\end{equation}

\noindent
while for $\Sigma_0 > \Sigma_1$

\begin{equation}
\Sigma(\alpha)= 
\left\{
\begin{array}{ll}
\frac{\Sigma_1(w_2-w_1)+(\Sigma_0(1+w_2)+\Sigma_1(w_1-w_2))\left(\frac{\alpha_0}{\alpha}\right)^{2w_2+2}}{1+w_1} & \mbox{, if $\alpha \leq \alpha_1$} \\
\Sigma_1\left(\frac{\alpha_1}{\alpha} \right)^{2w_1+2} & \mbox{, if $\alpha > \alpha_1$} \; .
\end{array}
\right.
\end{equation}

Another obvious choice is a polytrop EOS. 
Note, however, that in general a polytrop EOS is supposed to relate the rest energy density $\varepsilon$, 
as opposed to the total energy density $\sigma$, 
to the pressure in which case it takes the functional form $p=A\,\varepsilon^\kappa$, 
with parameters $A$ and $\kappa$. 
Thereby, whenever we replace $\varepsilon$ by $\sigma$ the `inverse form' of the polytrop EOS reads as (see, e.g., \cite{Read08}) 

\begin{equation}
\label{polytrop}
\Sigma(\Theta) = \left(-\frac{\Theta}{A}\right)^{1/\kappa}-\frac{\Theta}{\kappa-1} \;\, .
\end{equation}

\noindent
The solution of (\ref{general-eos}) can be given explicitly as 

\begin{equation}
\label{polytropsigma}
\Sigma(\alpha) = \left(-\frac{\Theta_0}{A}\right)^{1/\kappa} 
\left(\frac{\alpha_0}{\alpha}\right)^2 
- \frac{\Theta_0}{\kappa-1}\left(\frac{\alpha_0}{\alpha}\right)^{2\kappa}\,,
\end{equation}

\noindent
where $\Theta_0$ is related to the value of $\Sigma_0$ by (\ref{polytrop}). 

It is straightforward to check that the DEC is satisfied if and only if $\kappa\in(1,2]$. 
Assume, now that if we are given both of the values $\Sigma_0$ and $\Theta_0$ and the EOS is chosen 
to possess the form (\ref{polytrop}) then $\kappa$ has to take value in the interval $[1+w_0,2]$, where $w_0=-\Theta_0/\Sigma_0$. 
Then the square of initial speed of sound at $\Sigma_0$ can be given as $\CS^2|_{\Sigma_0}=\kappa w_0/(1+w_0)$.

In closing this subsection let us emphasize again that whenever an EOS,
$\Theta=\Theta(\Sigma)$, is chosen $\Sigma$ may be given as a function
$\Sigma=\Sigma(\alpha)$ and, in virtue of (\ref{V}), the effective potential can be given 
as an explicit function $V(\alpha;\eta)$ depending exclusively 
on the dimensionless radius $\alpha$ and the dimensionless de Sitter density parameter $\eta$.

\subsection{Gravastars in equilibrium}

In this subsection considerations will be restricted to the case of gravastars in equilibrium.
Equilibrium, as usual, signifies those configurations in rest for which forces are also balanced, i.e.
$V(\alpha_0; \eta)=0$ and $V'(\alpha_0; \eta)=0$,
where $\alpha_0$ stands for the value of $\alpha$ in equilibrium, 
while prime, $'$, denotes derivative with respect to $\alpha$. 
As is expected the vanishing of $V$ and $V'$ is equivalent to 
the vanishing of $\dot{\alpha}$ and $\ddot{\alpha}$, which can be verified by making use of (\ref{eqV}).
In depicting of the configuration space  of gravastars in equilibrium 
we shall apply the values $\Sigma_0 = \Sigma(\alpha_0)$ and $\Theta_0
= \Theta(\alpha_0)$, which, by making use of (\ref{sigma-dyn}) and
(\ref{theta-dyn}), along with the vanishing of $\dot{\alpha}$ and
$\ddot{\alpha}$ at $\alpha=\alpha_0$, can be given as

\begin{equation}
\label{sigma-stat}
\Sigma_0 = \frac{1}{8\pi\alpha_0}\left(\sqrt{1-\eta\alpha_0^2}
-\sqrt{1-\alpha_0^{-1}}\;\right) 
\end{equation}
and 
\begin{equation}
\label{theta-stat}
\Theta_0 =
\frac{1}{16\pi\alpha_0}\left(\frac{1-2\eta\alpha_0^2}{\sqrt{1-\eta\alpha_0^2}}
- \frac{1-\frac12\alpha_0^{-1}}{\sqrt{1-\alpha_0^{-1}}}\right) .
\end{equation}

\noindent
Using the algebraic equations of (\ref{sigma-stat}) and (\ref{theta-stat})
and the condition that $\Sigma_0(\eta,\alpha_0)>0$, it can be shown that for a gravastar in
equilibrium the surface tension, $\Theta_0(\eta,\alpha_0)$,  is always negative.
This then justifies that no exotic matter (with negative pressure) is required to form 
the shell of a gravastar in equilibrium.

Note that whenever an EOS is given only one
of the values of $\eta$ and  $\alpha_0$ can be chosen freely because,
in virtue of (\ref{sigma-stat}) and (\ref{theta-stat}), 
$w_0 = -\Theta_0/\Sigma_0$ has to be consistent with the EOS.
On the other hand whenever no definite choice is made for an EOS then the configuration
space of gravastars in equilibrium may be parameterized by $\eta$ and
$\alpha_0$. In this case the value of $w_0(\eta,\alpha_0)$ gets to be
determined by relations (\ref{sigma-stat}) and (\ref{theta-stat}).

In depicting the configuration space of gravastars in equilibrium it is useful to identify the region on the $(\alpha_0,\eta)$ plain 
where the DEC holds (see figure \ref{figure2}). In doing so recall
first that whenever $\Sigma_0$ is non-negative for a gravastar in equilibrium
$\Theta_0$ has to be negative. This immediately implies that the DEC reduces to $0\leq w_0\leq 1$; 
furthermore the upper boundary of the
subregion of the $(\alpha_0,\eta)$ plain where the DEC is satisfied is signified by
the $w_0=1$ curve. This curve---determined by the implicit relation $w_0(\eta,\alpha_0)=1$---can be given as 

\begin{equation}
\label{DEC}
\eta^{\rm DEC}_{\rm max}(\alpha_0) =
\frac{60\alpha_0^2-36\alpha_0-25+(5-6\alpha_0)
  \,\sqrt{100\alpha_0^2-124\alpha_0+25}}{128\,\alpha_0^3\,(\alpha_0-1)} \;\, ,
\end{equation}

\noindent
which intersects the $\alpha_0$-axis at the value $\alpha_0=25/24$ (see figure \ref{figure2}). 
Consequently, whenever the DEC holds
$\alpha_0$ has to be greater than or equal to $25/24$. 
This value of the minimal radius immediately imposes a strict limitation 
on the compactness, expressed by the $2m/r$ ratio, of a gravastar in equilibrium. 
It also follows from (\ref{DEC}) that for small $\eta$ the approximation $\eta^{\rm DEC}_{\rm max}\approx \frac25\alpha_0^{-3}$ holds.

For comparison it is useful to recall that the value of the ratio $2m/r$ is also restricted 
in the case of conventional star models composed by matter satisfying the DEC. 
In the case of spherically symmetric static
configurations under various assumptions concerning the energy density and
pressure of the star the ratio $2m/r$ has been investigated. A comprehensive summary of the pertinent results can be
found in  \cite{Karageorgis07}. It is shown there that if pressure is non-negative (although it may be anisotropic) 
and the DEC holds, then the numerical value of the sharp upper bound for $2m/r$ is $0.963$.\footnote{
This numerical value is claimed  to be precise up to three digits in \cite{Karageorgis07}.} It is
interesting that the corresponding upper bound for a
gravastar in equilibrium is a little bit smaller $24/25=0.96$. 
One of the readings of this finding is that whenever the DEC holds 
the radius of the most compressed gravastar in equilibrium is definitely larger than the radius of
the most compressed conventional star. This indicates then that instead of asking whether these gravastars can be distinguished from BHs,  
by any sort of astrophysical observations, it is more appropriate to ask whether they can be distinguished from the most compact stars.

\subsection{Stable equilibrium}

In deciding whether a gravastar is in equilibrium there has been no need to choose a specific EOS. 
Nevertheless, whenever we want to study the radial stability of these configurations we
shall need to know the EOS, at least, in a sufficiently small neighborhood of
$\Sigma_0$ as merely the knowledge of $\Theta_0$ and $\Sigma_0$ will not suffice then.

Before turning to the radial stability of gravastars there are some conceptual issues to be mentioned. 
In spite of the familiarity of the dynamical equation $\dot{\alpha}^2/2 + V(\alpha;\eta) = 0$ 
it is better not applying automatically the analogous one-dimensional classical mechanical arguments. 
The main reason is related to the fact that $V(\alpha;\eta)$ is not an external potential. 
In the case of the classical mechanical problem the `total energy' need not to be zero 
so we may simply either increase the kinetic energy term
or change the potential energy by moving out of our system from its
equilibrium. In both cases, since the potential is supposed to be intact, the
total energy is changed. As opposed to this in the case of a gravastar if the
initial velocity is changed the potential has also to be changed since the
formal `total energy' needs to remain zero. The change in the initial velocity
$\dot{\alpha}$ affects the initial surface energy density, $\Sigma_0$,
therefore the solution $\Sigma(\alpha)$ of (\ref{general-eos}), and, in turn,
$V(\alpha,\Sigma(\alpha); \eta)$ is also affected. 
This verifies that it is better to avoid thinking of $V(\alpha,\Sigma(\alpha); \eta)$ as an external potential. 

In carrying out the radial stability of gravastars we shall assume that $V$
smoothly depends on its indicated variables.  This implies then that
under a sufficiently  small perturbation of a gravastar in equilibrium the sign
of the  second derivative, $V''(\alpha_0)$, will be intact. 
In particular, since the perturbation increases
the kinetic term, the potential has to become negative in a sufficiently  small
neighborhood of $\alpha_0$ whence the yielded motion of the gravastar will be
bounded if the value of $V''(\alpha_0)$ is guaranteed to be positive. Accordingly, a
gravastar in equilibrium, with radius $\alpha_0$, is considered to be stable if
$V''(\alpha_0; \eta) > 0$. 

As was already mentioned, the vanishing of $V(\alpha_0)$ and $V'(\alpha_0)$---in the classical
mechanical analogy---corresponds to vanishing speed, $\dot{\alpha}=0$, 
and, in turn, the forces are balanced. The equilibrium
configurations can be classified according to the convex-concave character of the potential, 
i.e. it is reasonable to distinguish the following three categories.

\begin{figure}[ht]
\center
\vskip5.5mm
\includegraphics[width=89mm,angle=270]{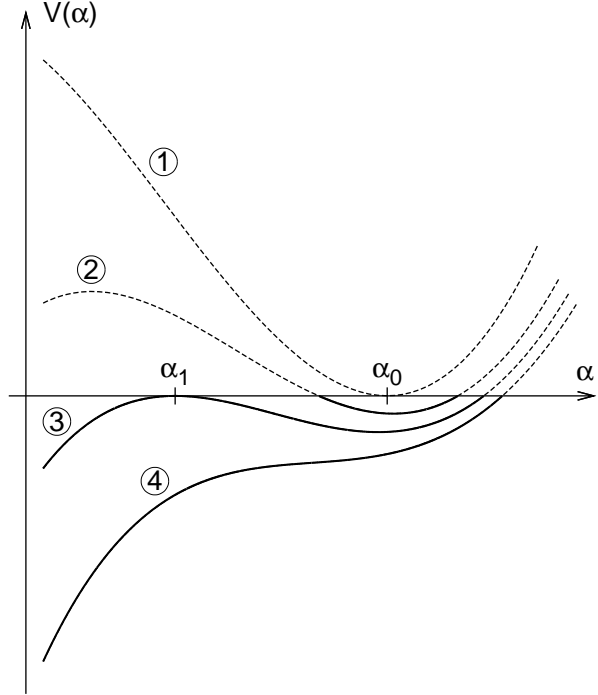}
\vspace{1mm}
\caption{
Typical potentials are indicated. 
Graph 1 depicts a potential relevant for stable equilibrium. 
The other three graphs, with labels 2, 3 and 4, represent the change of the potential 
caused by dropping more and more massive spherical dust shells onto the same initially stable gravastar. 
In particular, graphs 2 and 3 comes with a bounded motion of the resulted gravastar, 
while in the case of graph 3 the exactly maximal mass was dropped onto the initially stable configuration.
Graph 4 corresponds to the situation when the gravastar collapses to a BH.
}
\label{figure1}
\end{figure}

\begin{itemize}

\item[(a)] {\sl Stable}: A gravastar in equilibrium is considered to be stable
  whenever small perturbations of the system yield small local oscillations
  around the equilibrium state. In this case the potential has to be a convex function near $\alpha_0$.
  This may happen whenever $V''(\alpha_0)>0$. (See, e.g., the transition from potential 1 to potential 2 in
  figure \ref{figure1}). 

\item[(b)] {\sl Locally unstable}: This occurs if the potential is concave (at least from one side) near $\alpha_0$.
This means that small perturbations may yield non-local motions. 
This locally unstable situation may arise, e.g, whenever $V''(\alpha_0)<0$. (For an illustration see graph 3 in figure \ref{figure1}, 
where $\alpha_1$ takes the role of $\alpha_0$). 
If $V''$ is not continuous at $\alpha_0$ but has well-defined left- and right-hand-sided limit values the equilibrium state 
will be called locally unstable if either of these limit values is negative.

The vanishing of $V''(\alpha_0)$ may also occur. In this case the sign of the least non-zero  $\alpha$-derivative 
of $V$ determines the character of the potential. 
For instance, the graph of $V$ may change its character from concave into convex or vice versa through an inflexion point 
whenever $V'''$ is the first non-zero higher order derivative. 
This means that the equilibrium is locally unstable in one direction. 
However if the first non-zero $\alpha$-derivative of $V$ is $V''''$, then 
the graph of $V$ may have local minimum or maximum in accordance with the sign of $V''''$. 

\item[(c)] {\sl Neutral}: It may also happen that the potential is identically zero, $V(\alpha)\equiv 0$, 
in the neighborhood of $\alpha_0$.
In this particular case the equilibrium is called neutral (throughout this neighborhood). 

\end{itemize}

In determining the value of $V''(\alpha_0)$ recall first that $\CS^2$ can be given as 

\begin{equation}
\label{omega}
\CS^2 = -\frac{\rmd \Theta}{\rmd \Sigma} =
-\frac{\rmd\Theta/\rmd\alpha}{\rmd\Sigma/\rmd\alpha} \; .  
\end{equation}

\noindent
The terms $\rmd\Theta/\rmd\alpha$ and $\rmd\Sigma/\rmd\alpha$ can be expressed
by using equations (\ref{sigma-dyn}) and (\ref{theta-dyn}). 
For instance, in virtue of (\ref{V}), and since $\dot{\alpha}^2=-2V(\alpha)$ and $\ddot{\alpha}=-V'(\alpha)$, 
we find that $\rmd\Sigma/\rmd\alpha$ can be given
in terms of $V(\alpha)$ and $V'(\alpha)$ exclusively. Similarly,
$\rmd\Theta/\rmd\alpha$ can be seen to depend on $V(\alpha)$, $V'(\alpha)$ and
$V''(\alpha)$ exclusively, i.e. no higher order $\alpha$-derivatives of
$V(\alpha)$ are needed in evaluating $\rmd\Theta/\rmd\alpha$.  Therefore, in
the case of a gravastar in equilibrium, i.e. whenever $V(\alpha_0)=0$ and
$V'(\alpha_0)=0$, $V''(\alpha_0)$ can be expressed as a function of the
variables $\eta$, $\alpha_0$ and $\CSO^2=\CS^2\vert_{\alpha_0}$. 
This means---as was mentioned before---in carrying out the stability
analysis of gravastars in equilibrium only the derivative ${\rmd \Theta}/{\rmd \Sigma}=-\CS^2$ of
the EOS at $\Sigma_0$ comes into play.

It can be shown that for stable configurations the square of the speed of sound is bounded
from below by

\begin{equation}
\label{omegamin0}
c^2_{s,\mathrm{min}} =\left.-\frac{\partial\Theta/\partial\alpha}{\partial\Sigma/\partial\alpha}\right|_{\alpha_0}
=-\frac{\partial\Theta_0/\partial\alpha_0}{\partial\Sigma_0/\partial\alpha_0} \; ,
\end{equation}\

\noindent
where $c^2_{s,\mathrm{min}}$ is nothing but the square of the speed of sound $\CSO^2$ at $\alpha_0$
for configurations with $V''(\alpha_0) = 0$.
This, in particular, means that $V''(\alpha_0) > 0$ whenever $\CSO^2 > c^2_{s,\mathrm{min}}$.

To see that (\ref{omegamin0}) holds note first that $\Sigma$ and $\Theta$---given originally 
as functions $\Sigma(\alpha,\dot\alpha;\eta)$ and $\Theta(\alpha,\dot\alpha,\ddot\alpha;\eta)$---in virtue of the relations 
$\dot\alpha^2=-2V(\alpha;\eta)$ and $\ddot\alpha=-V'(\alpha;\eta)$ can be given 
as the functions of the form $\Sigma=\Sigma(\alpha,V(\alpha;\eta);\eta)$ and $\Theta=\Theta(\alpha,V(\alpha;\eta),V'(\alpha;\eta);\eta)$. 
Then (\ref{omega}), along with the relations $\rmd\Sigma/\rmd\alpha=\partial\Sigma/\partial\alpha +(\partial\Sigma/\partial V)\,V'$ 
and $\rmd\Theta/\rmd\alpha=\partial\Theta/\partial\alpha +(\partial\Theta/\partial V)\,V'+(\partial\Theta/\partial V')\,V''$, 
and the vanishing of $V'(\alpha;\eta)$ and $V''(\alpha;\eta)$ in the case of a gravastar in equilibrium justify the first equality in (\ref{omegamin0}). 
The second one holds because, in virtue of 
(\ref{sigma-dyn}) and (\ref{sigma-stat}) $({\partial\Sigma/\partial\alpha})\vert_{\alpha_0}=({\partial\Sigma_0/\partial\alpha_0})$, 
and similarly, in virtue of (\ref{theta-dyn}) and (\ref{theta-stat}) $({\partial\Theta/\partial\alpha})\vert_{\alpha_0}=({\partial\Theta_0/\partial\alpha_0})$. 
By taking into account (\ref{sigma-stat}) and (\ref{theta-stat}) the minimal value of  $\CSO^2$ can be given as

\begin{equation}
\label{omegamin}
\fl \hspace{2em}
c^2_{s,\mathrm{min}}(\eta,\alpha_0) = 
\frac{4\alpha_0^2(1-\alpha_0^{-1})^{3/2} -
  (3-6\alpha_0+4\alpha_0^2)(1-\eta\alpha_0^2)^{3/2}} 
{4(3-5\alpha_0+2\alpha_0^2)(1-\eta\alpha_0^2)^{3/2}
  +8\alpha_0^2(\alpha_0^2\eta-1)(1-\alpha_0^{-1})^{3/2}} \; .
\end{equation} 

\begin{figure}[h!]
\center
\includegraphics[width=100mm,angle=270]{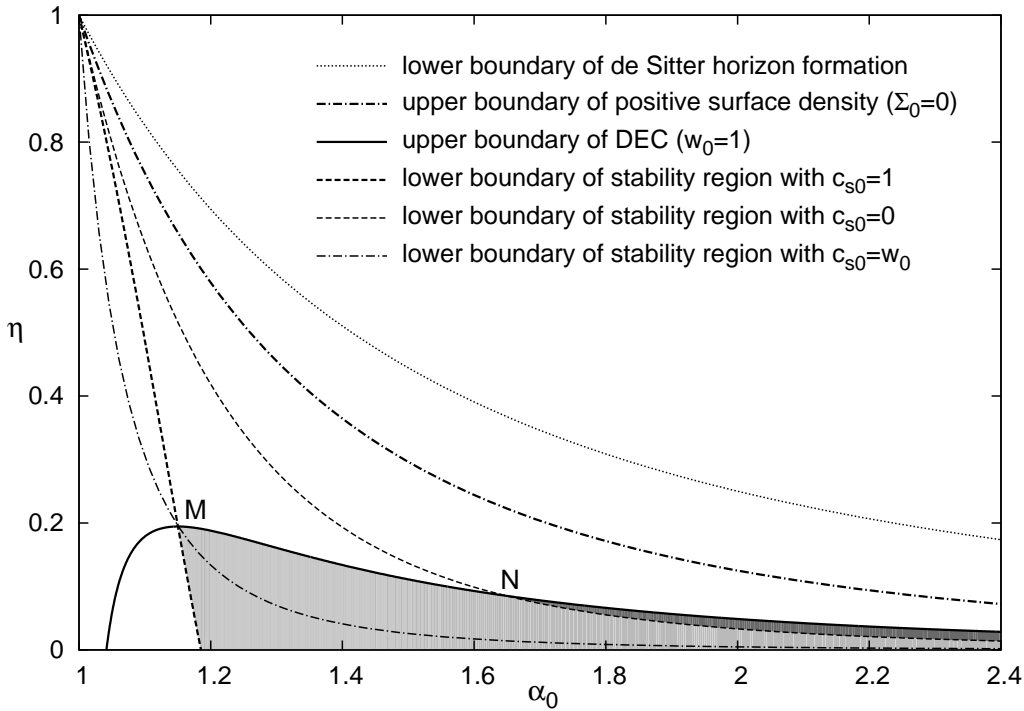}
\caption{
The configuration space of gravastars in equilibrium.  
The points below the uppermost dotted line represent configurations 
where no cosmological horizon may occur in the de Sitter region. 
At the points of the thick dotted line the surface energy density vanishes. 
Below that line the proper mass of the shell is always positive. 
The thick continuous line is determined by equation (\ref{DEC}) 
and it shows the upper boundary of the region where DEC holds.
This curve meets the horizontal axis at $\alpha_0 = 25/24$ 
and it has its global maximum at the point $M$ 
with coordinates $(\alpha_0,\eta)\approx(1.1503,0.1944)$. 
The dashed thick line represents configurations 
with $V''(\alpha_0)\equiv 0$ and with $\CSO^2=1$.
Stability holds, i.e. $V''>0$ with $\CS^2=1$, to the right of this curve, 
while instable configurations are to the left of this curve. 
The thin dashed line represents configurations 
with $V''(\alpha_0)\equiv 0$ and $\CSO^2=0$. 
This curve also separates stable and unstable configurations 
when $\CSO$ is restricted to zero.
The dashed thin line enters to the DEC region at the point $N$ 
with coordinates $(\alpha_0,\eta)\approx(1.6470,0.0855)$. 
On the thin dashed-dot line
$V''\equiv 0$ and $\CSO^2=-\Theta_0/\Sigma_0$ holds. 
The dashed thick line, the thin dashed-dot line 
and the curve determining the boundary of the DEC region all intersect at $M$.
Two subregions are indicated by gray shading. 
In the shaded region stable gravastars can be found 
within the domain where DEC holds and the speed of sound at $\alpha_0$ does not exceed the speed of light.  
The dark gray region represents stable gravastar configurations 
irrespective of the value of $\CSO^2$ until it is non-negative.
}
\label{figure2}
\end{figure}

In figure \ref{figure2} the configuration space of gravastars in equilibrium is illustrated 
with the help of the dimensionless parameters $\eta$ and $\alpha_0$. 
In explaining the meaning of some of the characteristic curves plotted in figure \ref{figure2} 
note that the thin dashed, dashed-dotted and thick dashed lines do correspond to configurations 
for which the square of the speed of sound $\CSO^2$ at $\alpha_0$ takes the minimal values $c^2_{s,\mathrm{min}}=0,w_0,1$, respectively. 
Correspondingly, the thick dashed line represents the lower bound of the radius for stable configurations for 
which the sound speed does not exceed the speed of light. 
If the DEC is also required to hold, then $\alpha_0$ has to be larger than $\approx 1.15028$ 
which is the $\alpha_0$ value of the configuration represented by the point $M$ in figure \ref{figure1}.

\begin{figure}[h!]
\vspace{-5mm}
\center
\includegraphics[width=110mm,angle=270]{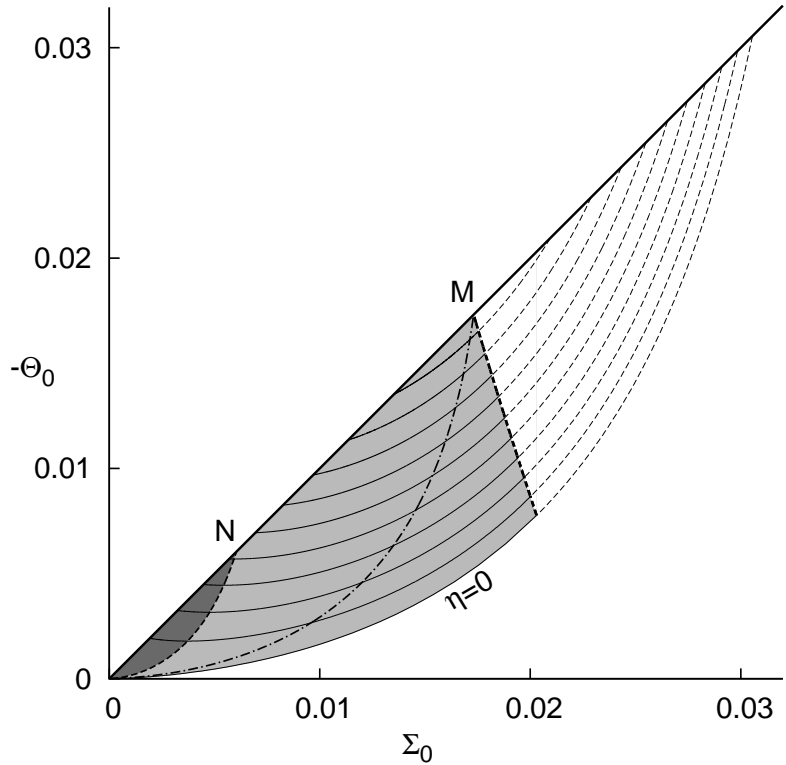}
\caption{
The configuration space of gravastars on the $(\Sigma_0,-\Theta_0)$ plane. 
An EOS may be represented by a curve on this plane. 
In order to provide a better orientation 
the $\eta=const$ curves for values ranging from $\eta=0$ to $\eta=0.18$, 
with separation $\Delta\eta=0.02$, are indicated. 
The styles of the sheading of the distinguished subregions 
and that of the characteristic curves and points are the same as in figure \ref{figure1}. 
}
\vspace{1mm}
\label{figure3}
\end{figure}

In figure \ref{figure3} a different representation of the very same configuration space, 
as depicted in figure \ref{figure2}, is given on the $(\Sigma_0,-\Theta_0)$ plane. 
The points representing gravastars in equilibrium by the values $(\eta,\alpha_0)$ in figure \ref{figure2} are mapped 
as $(\eta,\alpha_0) \mapsto \left(\Sigma_0(\eta,\alpha_0),\Theta_0(\eta,\alpha_0)\right)$. 
The styles of the sheading of the distinguished subregions 
and those of the characteristic curves and points are in correspondence in these two figures. 
An interesting feature of figure \ref{figure3} is that, in virtue of the above argument in connection with $c^2_{s,\mathrm{min}}$, 
the slope of the tangent at a point of an $\eta=const$ curve is exactly the value of $c^2_{s,\mathrm{min}}$ 
relevant for the configuration represented by the pertinent point.

\section{Dust shell falling onto the gravastar}
\label{Collision-section}

The interaction of a gravastar with a spherical dust shell falling onto it will be studied in this section. In advance to
the arrival of the dust shell onto the surface of the gravastar the history of the shell is completely determined by its
relativistic equation of motion on a Schwarzschild background. To make our
investigations somewhat realistic we shall assume that the collision of the
dust shell and the gravastar is totally inelastic, i.e. all the dust particles will
be assumed to move together with the surface layer of the yielded dynamical
gravastar. In determining its basic parameters---the velocity, mass
density, etc.---the conservation of the $4$-momenta
will be applied. Finally, by inspecting the new potential relevant for the composed system the dynamics 
of the yielded gravastar can be determined. 

\noindent
\begin{figure}[ht]
\vspace{-9mm}
\center
\includegraphics[height=77mm]{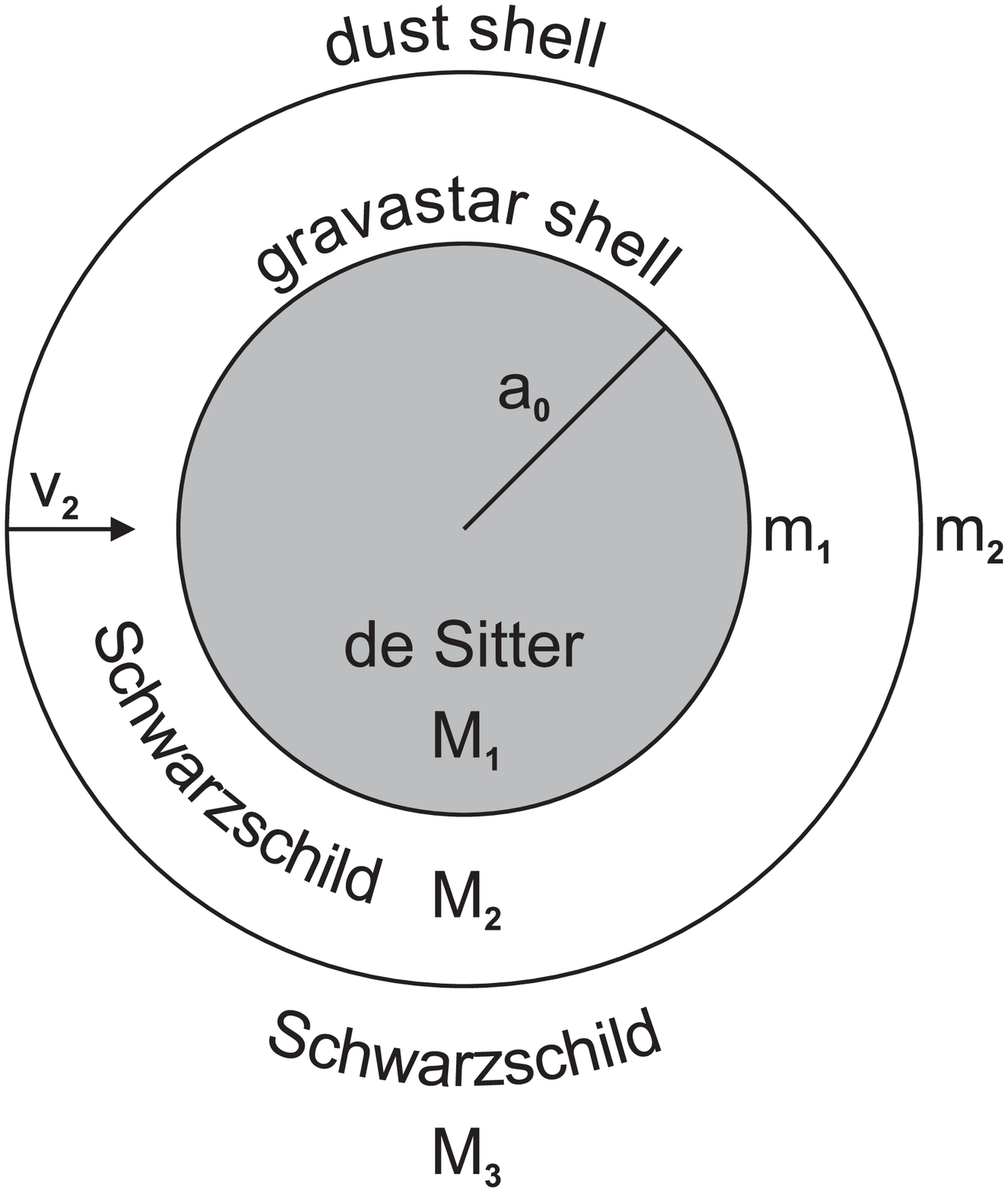}
\includegraphics[height=77mm]{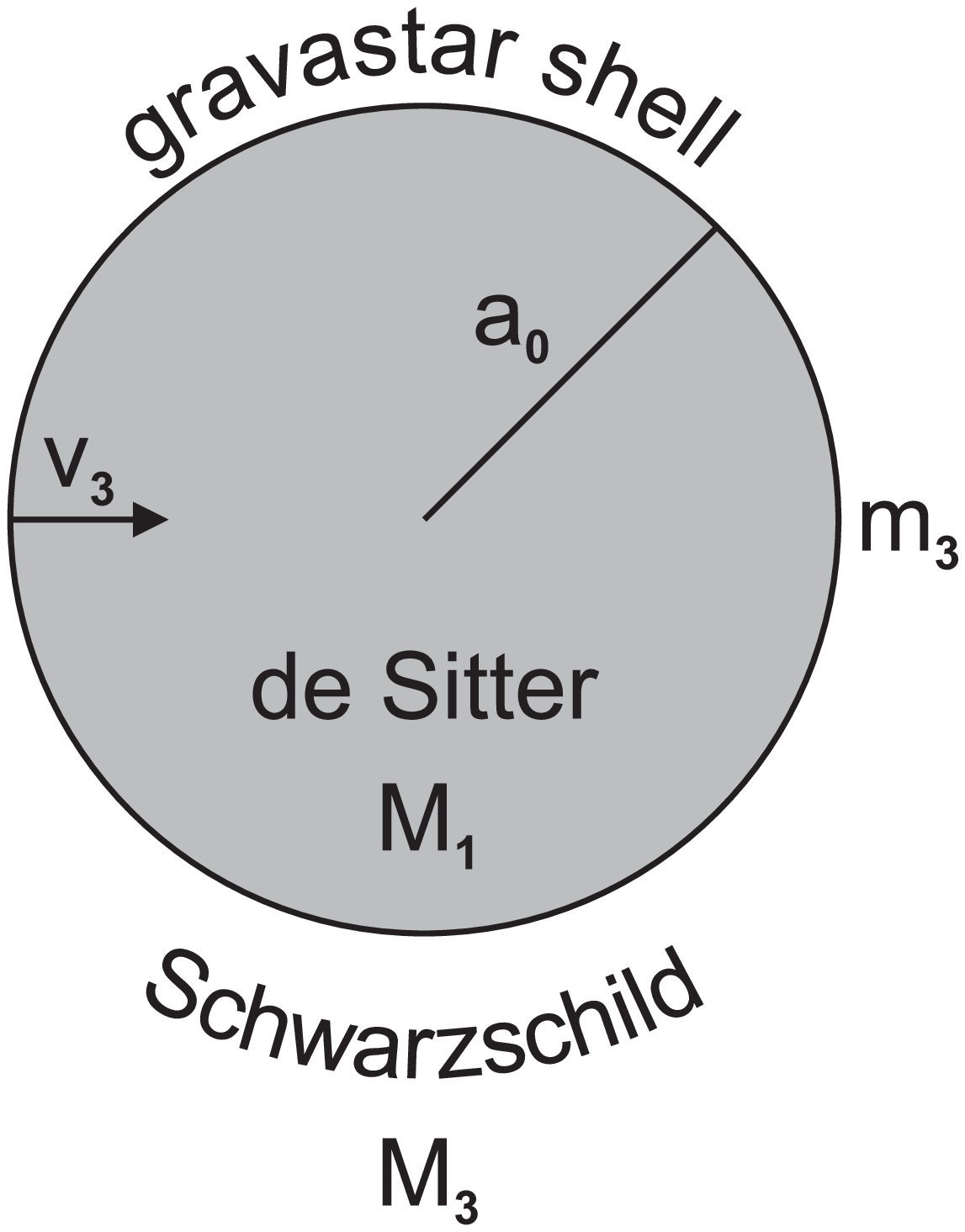}
\vspace{1.5mm}
\caption{
Left: the de Sitter interior with total mass $M_1$ inside the radius $a_0$,\
Schwarzschild exterior with mass parameter $M_2$ gravastar shell with rest mass $m_1$ 
and the moving dust shell with rest mass $m_2$ are shown. 
Right: the moment of the inelastic collision is indicated.}
\label{coll}
\end{figure}

According to the above-outlined program we shall start with a gravastar in equilibrium 
the total mass of which $M_2$ will be used as our reference unit. 
The dimensionless radius and the mass ratio parameter of this gravastar are $\alpha_0=a_0/(2M_2)$ and $\eta= 8 k M_2^2$, respectively.  
We assume that a dust shell, with gravitational mass $\varepsilon M_2$, falls
onto the surface of the gravastar and the collision between the dust shell and
the surface of the gravastar is totally inelastic. This, in particular, means that once
the  collision occurs a new three-layer dynamical gravastar
forms. Figure \ref{coll} provides a simple depiction of the underlying process. 
(For another schematic spacetime diagram see figure \ref{space-time-diagram}.)  
Denote the proper mass of the initial
shell of the gravastar, of the dust shell and of the shell of the gravastar after the
collision by $m_1$, $m_2$ and $m_3$, respectively. 
The total mass of the system is $M_3=(1+\varepsilon)M_2$.  
To guarantee $a_0$ to be greater than the Schwarzschild radius of the event
horizon of the yielded gravastar, we require the inequality $\varepsilon<\alpha_0-1$ to
hold. 

The dynamical characterization of the dust shell in advance to the collision can be given based on the results outlined in the appendix. 
Concerning the velocity of the particles of the dust shell at the moment of the collision 
it will be assumed that either it is equal to the velocity that would be acquired by the shell if it was starting to move toward the gravastar from rest at infinity, 
i.e. $v_2(\infty)=0$, or, in the second case, the particles of the dust shell will be assumed to be simply attached to the surface of the gravastar 
with zero velocity, i.e. $v_2(\alpha_0)=0$.\footnote{Although both of these assumptions correspond to certain limiting cases 
we expect that they provide some clue about more generic cases with an initial velocity between two of them.} 

Note that in the $v_2(\infty)=0$ case the gravitational mass of the dust shell equals to its rest mass, i.e. $m_2=\varepsilon M_2$ 
(see the appendix for more details), and also that, in virtue of (\ref{rdot}), 
the velocity $v_2=\rmd r/\rmd\tau_2$ at the moment of the collision can be given as

\begin{equation}
\label{v2}
v_2 = -\frac{\sqrt{16\alpha_0 + 8\varepsilon\alpha_0 + \varepsilon^2}}{4\alpha_0} \; .
\end{equation}

\subsection{Collision}

The conservation of the $4$-energy-momentum can be used to determine the
radial energy and the radial momenta of the surface of the gravastar yielded
by the collision.  The relevant equations are derived in the appendix where a
detailed and generic investigation of an inelastic collision of two concentric spherically symmetric matter shells is also
carried out. Because of spherical symmetry the radial component of the
4-momentum conservation does correspond to the relativistic 3-momentum
conservation---given by equation (\ref{appendixv3}). In virtue of this
relation the radial velocity, $v_3=\rmd a/\rmd\tau_3$, of the boundary of the
yielded gravastar if $v_2(\alpha_0)\not=0$ can be given as
\begin{equation}
\label{v3}
v_3 = \frac{8\pi\alpha_0 F(\eta,\alpha_0,\varepsilon)}{\Sigma_3} \; , \\
\end{equation}
where
\begin{equation}
\label{F}
F(\eta,\alpha_0,\varepsilon) =  -\frac{\Sigma_0(\eta,\alpha_0)^2
\,(1-1/\alpha_0)\,(\varepsilon/2)\,\sqrt{16\alpha_0 + 8\varepsilon\alpha_0 +
  \varepsilon^2}}{1-2\,\eta\,\alpha_0^3+\eta^2\alpha_0^6
  -4096\,\pi^4\alpha_0^6\,\Sigma_0(\eta,\alpha_0)^4} \; . \\
\end{equation}

\noindent
Here $\Sigma_0(\eta,\alpha_0)$ denotes the energy density of the original
gravastar in equilibrium given by (\ref{sigma-stat}).
 
The temporal component of the 4-momentum conservation  corresponds to the
energy conservation  which can be seen to be equivalent to the dynamical
master equation of gravastars (\ref{sigma-dyn}).  Using the parameters of the
newly formed gravastar the master equation can be given in terms of the
above-introduced dimensionless parameters as
\begin{equation}
\label{Sigma3}
\Sigma_3 = \frac{\sqrt{1-\eta\alpha_0^2+v_3^2}
  -\sqrt{1-(1+\varepsilon)/\alpha_0+v_3^2}}{8\pi\alpha_0} \; . 
\end{equation}
The substitution of (\ref{v3}) into (\ref{Sigma3}) then yields the following
equation:
\begin{equation}
\label{eq}
\Sigma_3 = \sqrt{A(\eta,\alpha_0)^2
  +\frac{F(\eta,\alpha_0,\varepsilon)^2}{\Sigma_3^2}} 
-\sqrt{B(\alpha_0,\varepsilon)^2+\frac{F(\eta,\alpha_0,\varepsilon)^2}{\Sigma_3^2}} \; ,
\end{equation}
with
\begin{eqnarray}
A(\eta,\alpha_0)  =  \frac{\sqrt{1-\eta\alpha_0^2}}{8\pi\alpha_0}\,,\quad \textrm{and} \quad
B(\alpha_0,\varepsilon)  =  \frac{\sqrt{1-(1+\varepsilon)/\alpha_0}}{8\pi\alpha_0} \;\, .
\end{eqnarray}

\noindent
Equation (\ref{eq}) is an algebraic relation for $\Sigma_3$ possessing the
only positive root
\begin{equation}
\label{root}
\Sigma_3(\eta,\alpha_0,\varepsilon) = \sqrt{A^2+B^2-2\sqrt{A^2 B^2 + F^2 }} \; ,
\end{equation}
where $\Sigma_3$ denotes the energy density of the boundary of the newly formed gravastar. Note that the value of $\Sigma_3$ is independent from the EOS.

Whenever $v_2(\alpha_0)=0$ the value of $\Sigma_3$ can be determined by (\ref{Sigma3}) where the substitution $v_3=0$ has to be used.

\subsection{After the collision}

As discussed above once the collision occurs $\Sigma_3$ and $v_3$ can be determined. 
Note that to be able to determine what happens after the collision we also need to know the
EOS of the newly formed surface. Without this information we
cannot determine the new surface energy density as a function of the radius, which is essential in assigning the
new effective potential.

Concerning the new EOS, since the pressure of the falling dust shell is zero,
it is tempting to assume  that the surface tension remains intact during the
process of collision. This idea could be supported by a simple-minded
summation of partial pressures.  However, if this happened the particles of
the dust should have stopped at the surface of the gravastar  without
interacting with the particles forming the surface of the gravastar. This
appears to be baseless, and more importantly, to be incompatible with the
physical picture of inelastic collisions. To resolve the associated
discrepancy, hereafter we shall assume that instead of the surface tension the EOS of the boundary remains
intact during the inelastic collision. This means that the dust shell `melts' into the matter of the surface of the gravastar.

In determining the new potential, as previously, (\ref{V}) may be applied  
since after the collision a modified three-layer gravastar is formed. 
The only distinction is that the new dynamical 
variables---indicated by tilde---have to be used in (\ref{V}). 
For the new gravastar $\widetilde{\eta}=(1+\varepsilon)^2\eta$ holds as the mass of the outer Schwarzschild region, 
used as a reference unit, changes from $M_2$ to $M_3=(1+\varepsilon)M_2$.
Accordingly, the dimensionless surface density and the dimensionless radius change as
$\widetilde{\alpha}=\alpha/(1+\varepsilon)$ and $\widetilde{\Sigma}=(1+\varepsilon)\,\Sigma$.
Combining all these observations the new potential may be given as  

\begin{equation}
\label{newV}
\widetilde{V}(\alpha;\eta) = V((1+\varepsilon)^2\eta,(1+\varepsilon)\Sigma_{\mathrm{new}}(\alpha),\alpha/(1+\varepsilon)).
\end{equation}

\noindent
where $\Sigma_{\mathrm{new}}(\alpha)$ is the solution of 
(\ref{general-eos}) with the initial condition 
$\Sigma_{\mathrm{new}}(\alpha_0)=\Sigma_3$ provided that the EOS remains intact as assumed.

\section{Results}
\label{Results}

By inspection of the new effective potential the dynamics of the gravastar can be determined. 
If the new potential possesses the form of graph 2 of figure \ref{figure1} 
the motion of the gravastar surface remains bounded and we can determine the minimum and maximum value of the radius. 
In particular, the maximum mass which may be dropped onto the original gravastar without converting it into a BH can also be determined. 

\subsection{The applied method}

First note that after collision the initial velocity points inward (unless $v_2(\alpha_0)=0$);
therefore, the investigation has to start by deriving whether the new effective potential 
has any root in the interval $\alpha\in[1,\alpha_0]$.
If not then an event horizon will form and the gravastar collapses to a BH.
The examination of the potential is made by a numerical maximum finding algorithm. 
If the maximum is found to be less than zero, the collapse to a BH will occur.
Using an interval bisection method,
changing the initial gravitational mass of the dust shell to be dropped,
the numerical value of the maximal gravitational mass, $\EMAX\cdot M_2$,
which may fall onto the gravastar without converting it into a BH, can be determined.
Whenever $v_2(\alpha_0)\not=0$ and there is a root in the interval $\alpha\in[1,\alpha_0]$,
the motion is bounded and it happens so that $\alpha_1\leq \alpha\leq \alpha_2$, 
where $\alpha_1$ is the largest root that is smaller than $\alpha_0$ and 
$\alpha_2$ is the smallest root that is larger than $\alpha_0$. 
If $v_2(\alpha_0)=0$ then $\alpha_0$ coincides with either the upper or lower bound of the motion.

It is an inherent property of the gravastar model that we cannot have information about the de Sitter energy density inside the gravastar, $\eta$, 
nor about the EOS of the shell before and after the collision. To have some quantitative results suitable hypotheses have to be applied. 
In order to avoid the use of a flood of parameters---each of 
which has to possess definite values in numerical investigations---, as indicated 
in subsection \ref{EOSsubsec}, in this paper we analyze three types of EOS: homogeneous linear, broken linear and polytrop. 

If we fix an EOS with all of its parameters,
then, as a function of $\eta$, the stable equilibrium radius, $\alpha_0$, is to be determined.
Note, that this radius can be determined by the intersection
of the specific EOS curve and the corresponding $\eta$-level curve in the figure \ref{figure3}.
Next we have to make an assumption about the EOS after the collision.
Since we have no observational clues, as it was indicated above, we shall assume that the EOS remains intact, 
which means that the dust shell melts into the matter of the gravastar shell yielding an increase of the pressure according to its mass contribution.
Note that in the case of an initially homogeneous linear EOS the possibility of keeping the surface tension to be intact will also be investigated. 
This latter case might be a good approximation whenever the contribution of the dust shell to the new shell's composition is negligible.

\subsection{Homogeneous linear EOS}

\begin{figure}[h!]
\center
\includegraphics[width=100mm,angle=270]{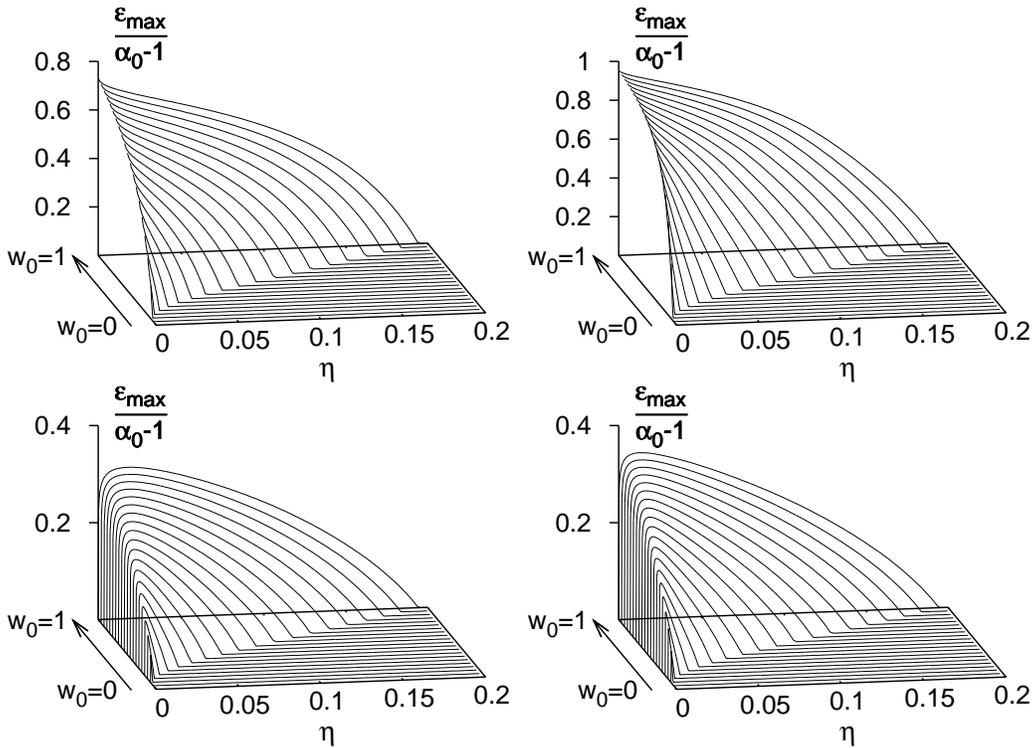}
\caption{
The normalized value $\EMAX/(\alpha_0-1)$ is shown 
as a function of $\eta$ and $w_0$ for the homogeneous linear EOS, 
where $\alpha_0$ stands for the maximal dimensionless radius allowed by the DEC. 
On the top two panels $v_2(\infty)=0$, while on the bottom two ones $v_2(\alpha_0)=0$. 
The panels to the left depict the intact EOS cases, 
while in the right ones the surface tension was kept to be intact.
}
\label{w0surf}
\end{figure}

As was mentioned in subsection \ref{EOSsubsec}, 
the simplest possible EOS is the homogeneous linear one. 
If the value of $\eta$ and $w_0$ are given that of $\alpha_0$ 
is determined by the implicit relation $w_0=-\Theta_0(\eta,\alpha_0)/\Sigma_0(\eta,\alpha_0)$. 
It is clearly visible in figure \ref{figure3} that for a given value of $w_0\,(=c_{s}^2\in[0,1])$ 
the $w_0\,\Sigma_0+\Theta_0=0$ line intersects any $\eta$ constant curve at most at two points. 
One of these intersections corresponds to a stable configuration, 
with the smaller $\Sigma_0$ value, while the other is unstable. 
Accordingly, once the choice for $\eta$ and $w_0$ is made first we determine the value of $\alpha_0$ by numerically solving the implicit relation $w_0=-\Theta_0(\eta,\alpha_0)/\Sigma_0(\eta,\alpha_0)$. 
Then, for any choice of the pair $(\eta,w_0)$, we determine the stable gravastar configuration, and as it was outlined above, 
the maximal mass of the shell that can be dropped onto it without converting it into a BH can also be determined. 
The results of the corresponding investigations are depicted by figure \ref{w0surf}. 
It is worthy to mentioned that some of the special cases depicted on these panels play distinguished role in various arguments. 
For instance, the case $w_0=1/3$ corresponds to ultrarelativistic gas, 
which---based on certain dimension reduction arguments that might not be straightforward 
to be applied in the present case---is represented in \cite{Zlosh} by a system with $w_0=1/2$.

Note that to have a stable configuration the value of $w_0=c_s^2$ has to be larger than $c^2_{s,\rm{min}}$ given by (\ref{omegamin}). 
In particular, $w_0=c_s^2$ may take all the values from the interval $[0,1]$ only for $\eta=0$, i.e. whenever the inner region is vacuum. 
Also note that $\EMAX$ has to be smaller than $\alpha_0-1$.
If $\varepsilon \geq \alpha_0-1$ occurred, 
the system composed of the gravastar and the dust shell would get beyond its own Schwarzschild radius before the collision occurred. 
This makes the use of the rescalled quantity $\EMAX/(\alpha_0-1)$ to be advantageous. 

\begin{figure}[h!]
\center
\includegraphics[width=100mm,angle=270]{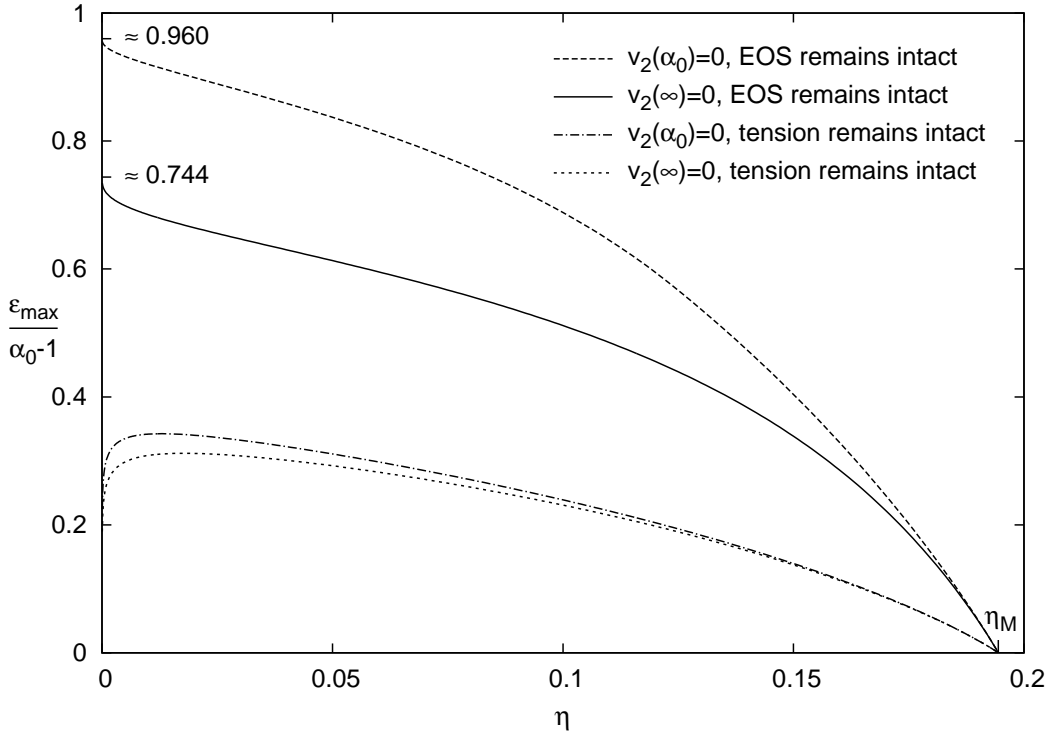}
\caption{
The $w_0=1$ sections of the four panels of figure \ref{w0surf} are shown 
providing a better comparison of $\EMAX$ for the considered configuration.
The order of the curves from top to bottom 
coincides with that of the legend at the upper-right corner.
}
\label{v2p2}
\end{figure}

Before interpreting figures \ref{w0surf} and \ref{v2p2} let us recall that a homogeneous linear EOS, 
$\Theta_0=-w_0\,\Sigma_0$, has $w_0$ as the only parameter. 
Therefore, it is also straightforward to investigate the case 
when both the surface tension and the functional form of the EOS remain intact during the collision. 
It is done by determining the value of the new $\widetilde w_0$ via the relation  $\widetilde w_0=-\Theta_0/\Sigma_3$. 
If this happens the particles of the dust shell do not adopt the properties of the particles comprising the surface of the gravastar 
as only the rest mass of the surface will be increased by the collision. In figures \ref{w0surf} and \ref{v2p2} four different subcases are considered. 
These subcases are yielded, on the one hand, by keeping the EOS or the surface tension to be intact and, on the other hand, 
by choosing the velocity of the particles of the dust shell at the moment of the collision to be such that either $v_2(\infty)=0$ or $v_2(\alpha_0)=0$, as discussed in section \ref{Collision-section}.

\begin{figure}[h!]
\center
\vspace{0.5mm}
\includegraphics[width=100mm,angle=270]{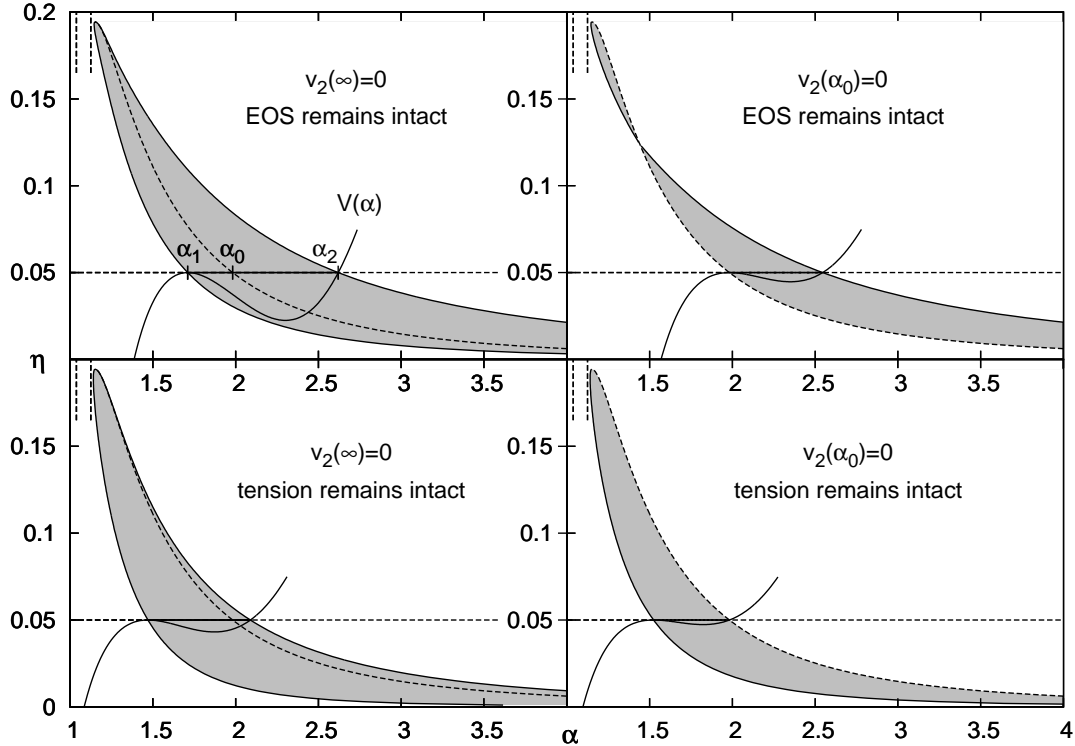}
\vspace{1.5mm}
\caption{
The dynamical range of the radius for maximally loaded gravastars is shown 
referring to the four basic configurations with $w_0=1$ considered in figure \ref{v2p2}. 
The $\eta=const$ sections of the gray regions represent the dynamical range 
of the radius relevant for the considered four subcases. 
The $\eta$ dependence of $\alpha_0$ of the gravastars in advance to the collision is indicated by the dashed curves. 
Whenever the initial velocity of the dust shell is nonzero (e.g., $v_2(\infty)=0$) 
the dashed line is in the interior of the grey region, 
whereas whenever  $v_2(\alpha_0)=0$---see, e.g., the right panels---the gravastar stars 
its motion at the edge (minimal or maximal $\alpha$ value) of the dynamical range. 
For the particular choice $\eta=0.05$ the corresponding potentials, are also plotted, 
where the zeros of the potentials are indicated by the dotted horizontal lines 
and the applied scales are the same on each of these panels.
The dotted vertical line with the larger $\alpha=9/8$ value correspond to the Buchdahl limit, 
while the other one with $\alpha\approx 1.0384$
indicates the corresponding lower bound to the size of most compact conventional stars satisfying the DEC.
}
\label{bananas}
\end{figure}

Figure \ref{v2p2} depict only the $w_0=1$---referred to as stiff matter in \cite{Visser04}---sections 
of the four cases indicated in figure \ref{w0surf} 
which offers a better comparison of the corresponding four different dynamical subcases. 
Figure \ref{bananas} does also refer to these four basic configurations with $w_0=1$. 
The horizontal sections, corresponding to an $\eta=const$ value, 
of the grey regions indicate the dynamical range of the  radius of gravastars. 
For the particular value $\eta=0.05$ the pertinent new potentials are also indicated. 

There are two interesting points to be mentioned that are indicated in figure \ref{bananas}. 
First, there are two dotted vertical lines on each of the panels. 
The one with the larger $\alpha$ value corresponds to the Buchdahl limit with $\alpha=9/8$ \cite{Buchdahl59}, 
while the other with a smaller $\alpha$ value indicates the corresponding lower bound, 
with $\alpha\approx 1.0384$, to the size of most compact conventional stars. 
On these panels of figure \ref{bananas} it is also visible that dynamical gravastars cannot get below the Buchdahl limit 
so they definitely remains less compact than the most compact conventional stars satisfying the DEC.

Second, on the right two panels with $v_2(\alpha_0)=0$ the grey regions on one side 
are bounded by the dashed lines representing the $\eta$ dependence of $\alpha_0$ of the initial gravastar. 
This is, on one hand, in accordance with intuition as the particles of the dust shell are simply simultaneously placed on the surface of the gravastar. 
What might be unexpected, on the other hand, is the following. 
Whenever the EOS remains intact (see the right-top panel of figure \ref{bananas}) 
there is a subregion where the dashed line bounds from above and another 
where it bounds from below.\footnote{The response of the initial gravastar 
for the placement of the dust shell onto its surface is as follows: 
By increasing the value of $\varepsilon$ from zero to $\EMAX$ first the radius increases. 
However, for configurations with $\alpha_0<1.44$ there is a turning point in $\alpha$  
and before reaching $\EMAX$  the radius is smaller than the initial $\alpha_0$.}
These subregions are separated by a point $(\eta,\alpha_0 )\approx (0.124,1.44)$ 
where the initial gravastar remains apparently at rest for $\varepsilon=\EMAX$, while whenever the
surface tension remains intact---as it is indicated by the bottom-right panel of figure \ref{bananas}---
the dynamical range for $\varepsilon=\EMAX$ is always bounded from above by the initial gravastar configurations.

\subsection{Broken linear EOS}

\begin{figure}[h!]
\center
\vspace{1.5mm}
\includegraphics[width=100mm,angle=270]{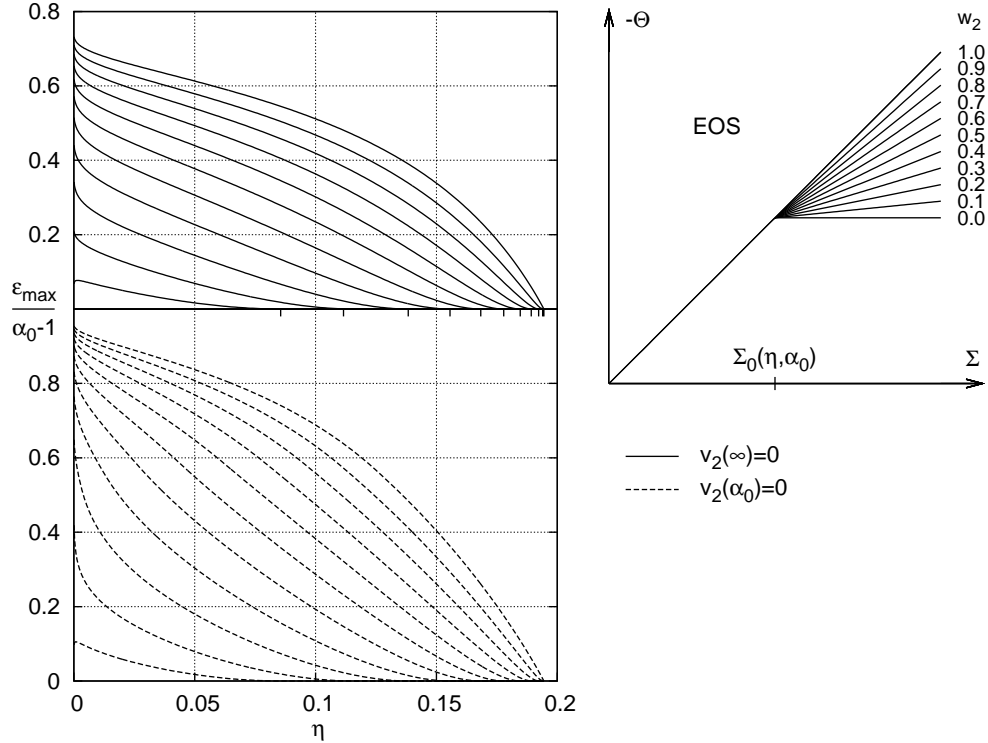}
\caption{
The applied broken linear EOS is shown on the top-right panel. 
$\alpha_0$ is chosen to be maximal such that the DEC is still guaranteed to hold. 
On the graphs of the left panels the value of $w_2$ decreases form 1 to zero, 
in steps $0.1$, while moving downward.
}
\label{w1dec}
\end{figure}

Since the gravastar radius has always to decrease at the moment of the collision the surface density must increase at this moment. 
Thus, the functional form of the new effective potential depends only on that of the EOS relevant for mass density greater than $\Sigma_0$. 
Motivated by this observation we have used the following particular broken linear EOS. 
$\Sigma_0(\eta,\alpha_0)$ was chosen to be the breaking point, 
i.e. the choice $\Sigma_1=\Sigma_0$ was made, and only the slope was varied for $\Sigma>\Sigma_0$. 
Since we have found that the higher the value of $w_0$ the more stable is a gravastar, 
in the broken linear EOS we have chosen $w_1=1$ and $w_2$ to be varied form zero to 1, in steps $0.1$, 
as it is indicated on the right panel of figure \ref{w1dec}.  

In figure \ref{w1dec} the $\eta$ dependence of $\EMAX$, 
i.e the maximal value of the gravitational mass of the shell that may be dropped onto a stable gravastar 
without converting it into a BH is shown for configurations with $v_2(\infty)=0$ and $v_2(\alpha_0)=0$. 
On both of the left panels the value of $w_2$ decreases from 1 to zero while moving from the top to the bottom. 
It is also visible that for each particular value of $w_2$ there is a maximal value of $\eta$ 
that is indicated by the vertical ticks on the horizontal axis of the the top-left panel. 
These figures do also justify---for the pertinent linear approximation of the EOS---that 
for given $\eta$ and $\alpha_0$ the larger the value of $c_{s0}=w_2$ 
the more stable the gravastar becomes in the sense that 
the allowed value of $\EMAX$ increases together with $c_{s0}$.

\subsection{Polytrop EOS}

\begin{figure}[h!]
\center
\vspace{1.5mm}
\includegraphics[width=70mm,angle=270]{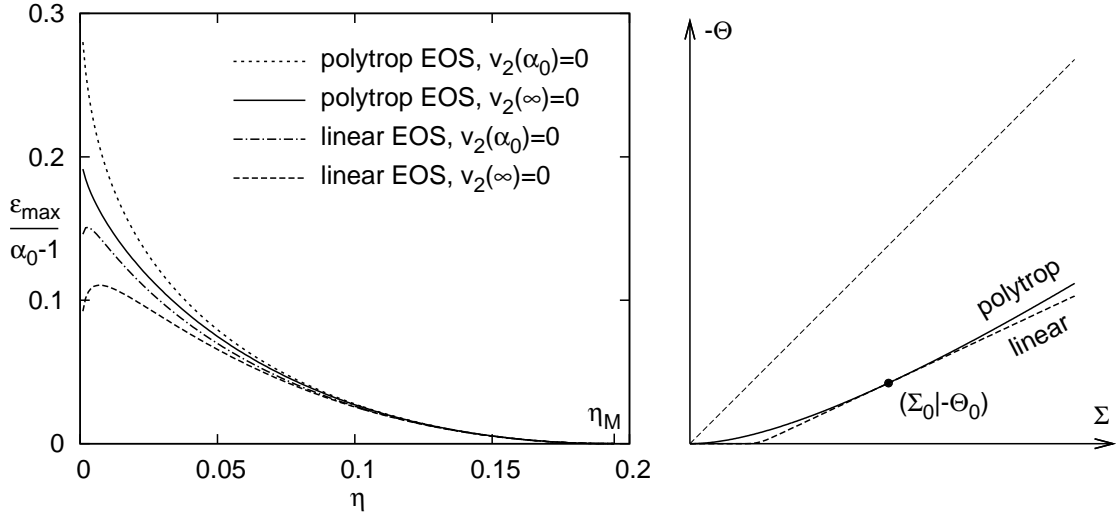}
\caption{
Right: the applied polytrop EOS with $\kappa=2$ and the linear one, 
having the same slope at $\Sigma_0$ are illustrated.
Left: for any choice of $\eta$ the corresponding value of $\alpha_0$ 
is determined by the intersection of the $\eta=const$ line and the dashed-dotted curve in figure \ref{figure2}. 
The $\eta$ dependence of $\EMAX/(\alpha_0-1)$ is shown 
for two types of EOS and for the two types of initial velocities, 
$v_2(\infty)=0$ and $v_2(\alpha_0)=0$. 
The order of the curves from top to bottom coincides 
with that of the legend at the upper right corner.
}
\label{kappa2}
\end{figure}

A polytrop EOS of the form of (\ref{polytrop}) has two parameters, $A$ and $\kappa$. 
Given a point $(\Sigma_0,\Theta_0)$---with $0<|\Theta_0|< \Sigma_0$---it is not obvious 
to choose at all the parameters $A$ and $\kappa$ 
to get an EOS fitting to this point except for the choice $\kappa=2$. 
To overcome this slight technical difficulty 
and also because for a given value of $A$ the choice $\kappa=2$ always yields the most stable configuration, 
we have chosen $\kappa=2$ and varied only the value of $A$. 
It is also informative to compare the behavior of a gravastar with a polytrop EOS 
and with a linear one having the same slope at $\Sigma_0$ as it is indicated 
by the nested panel in figure \ref{kappa2} relevant for the particular choice $\kappa=2$. 
For any choice of $\eta$ the value of $\alpha_0$ is chosen to be determined 
by the intersection of the $\eta=const$ line and the dashed-dotted curve in figure \ref{figure2}. 
Note that if $\alpha_0$ was chosen to be the maximal value for a given $\eta$ such that 
the DEC holds, the polytrop EOS would degenerate into the homogeneous linear one with $w_0=1$.

The $\eta$ dependence of the maximum of the normalized gravitational mass of the dust shell, $\EMAX/(\alpha_0-1)$,
---where $\EMAX$ is the maximal mass that may be dropped onto a gravastar without converting it into a BH---
is shown in figure \ref{kappa2} for two types of EOS and for the two types of choices concerning the initial velocities, 
$v_2(\infty)=0$ and $v_2(\alpha_0)=0$.
As it might have been anticipated, figure \ref{kappa2} justifies that a gravastar characterized 
by a polytrop EOS is more stable than the corresponding one with the linear EOS in consequence of the fact 
that the slope of the polytrop EOS is larger than that of the linear one for $\Sigma>\Sigma_0$.

\section{Conclusions}
\label{Conclusions}

In this paper, besides providing a short survey of gravastars in equilibrium, the radial stability of gravastars was investigated.
The latter was done by determining the response of a gravastar to the arrival of a dust shell onto its surface. 
Concerning the matter model of the surface of the gravastar three different types of EOS were applied. 
We also assumed that the DEC holds and the speed of sound $c_s$ satisfies the relations $0\leq c_s\leq 1$. 

Our most important findings are as follows. 
Among the investigated EOSs the homogeneous linear one with $c_s=1$ appears to provide the largest possible stability for gravastars 
in the sense that for a chosen $\eta$ the value of $\EMAX$ attains its largest possible value for  $c_s=1$. 
It would be interesting to find an analytical justification of this observation. 
It also follows from our investigations that once an EOS is fixed, for a given value of $\alpha_0$, the smaller the value of $\eta$, or alternatively, for a given value of $\eta$, the larger the value of $\alpha_0$ the more stable is a gravastar. The maximal mass of the shell that may be dropped onto a gravastar in equilibrium without converting it into a BH is also determined. The normalized value of this gravitational mass $\EMAX/(\alpha_0-1)$ was found to be the largest $\EMAX/(\alpha_0-1)\approx 0.96$  whenever the matter comprising the surface of the gravastar possesses a homogeneous linear EOS with $w_0=1$ and $v_2(\alpha_0)=0$. Correspondingly the value of $\EMAX\approx 0.96\cdot({a-2M})/{2M}$ is of the order 1 for a gravastar with radius $a\approx 3M$ which indicates that gravastars may be significantly more stable (although not smaller) than ordinary stars.

In the appendix a generic framework describing the totally inelastic collision of two shells is also shown. 
The reason behind considering inelastic collisions is that the de Sitter region of a gravastar is unstable against hydrodynamical excitations. 
Thereby, it is reasonable to assume that the particles of the dust shell merge to the particles of the surface of the gravastar. 
In most of the considered cases, for the sake of simplicity, we also assume that the EOS of the surface of the gravastar remains intact during the collision. 
All these assumptions indicate that there might be various generalizations of the simple model we applied. Nevertheless, our findings in all the investigated cases clearly manifest that even a dynamical gravastar cannot be more compact than the smallest possible ordinary stars of the same mass provided that for both types of models the DEC is guaranteed to hold.
\footnote{
It has been long-known that a gravastar---neither in the shell based model nor in its continuum correspondence---can be of considerably smaller size unless the DEC is violated \cite{Cattoen05,Lobo06}.}
This means that the main issue is not that whether gravastars can be distinguished from a BH. Assuming that they exist in reality it is more appropriate to ask whether they can be distinguished from compact regular stars.

Let us finally turn the case of a generic BH mimicking gravastar without restricting our considerations by referring to the DEC or to any particular EOS of the surface layer. Note that in this case the size of the gravastar is not bounded from below so, in particular, the assumption $\alpha_0\approx 1$ corresponding to the requirement of Mazur and Mottola may also hold. Now assume that the mass of this gravastar is $M$ and that the gravitational mass of the dust shell---falling onto it---is $\varepsilon\cdot M$. For simplicity we assume that the particles of the dust shell move toward the gravastar with velocity as if they started from rest at infinity, i.e. $v_2(\infty)=0$. Then, the mass of the composed system is $(1+\varepsilon) M$. In order to avoid the situation in which the composed system gets beyond its own Schwarzschild radius $R_S=2(1+\varepsilon) M$, in advance to the collision, for the radius of the initial gravastar, $R_{\alpha_0}=2 \alpha_0 M$, the inequality $R_S<R_{\alpha_0}$ must hold. This immediately implies then that $\varepsilon< \alpha_0-1$, and in turn that the closer the radius of the initial gravastar is to the Schwarzschild radius the smaller the mass of the dust shell, which may be dropped onto it without converting it into a BH, must be. In other words, the better is  a gravastar in mimicking a BH the easier is to destroy it by the described process. Note that since this argument does not refer to the internal structure of the gravastar model it applies not only to the simplified three-layer model of Visser and Wiltshire or to the original model of Mazur and Mottola but essentially to any analogous construction. 

\section*{Acknowledgments} 
This research was supported in part by OTKA grant K67942.

\appendix
\section*{Appendix}

\renewcommand{\theequation}{A.\arabic{equation}} 
\setcounter{equation}{0}
\setcounter{figure}{9}

This appendix is to introduce the basic equations governing the dynamics of a
totally inelastic collision of a pair of concentric spherical shells, the
outer of which is moving in a vacuum spacetime region. These equations are derived
by making use of the conservation of 4-momentum. In determining them
we have applied a suitable adaptation of the method of Nakao, Ida
and Sugiura  which was originally worked out in \cite{Nakao99} to describe the
conservation of 4-momentum in the special case of totally transparent shell crossings. In
what follows the abstract index notation will be applied. In particular, the
lowercase Latin indices will indicate the type of tensors while Greek indices
will refer to the components of tensors with respect to some specified basis
fields.

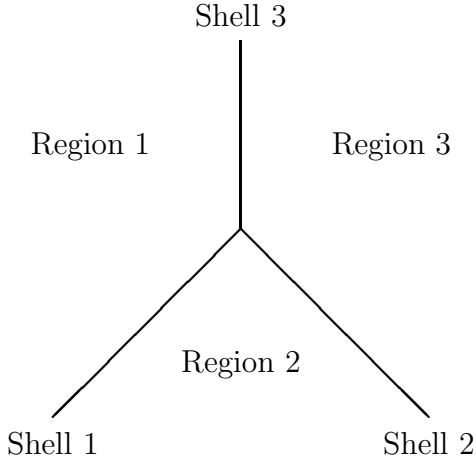
\begin{figure}[ht]
\vspace{2mm}
\center \unitlength 1mm
\begin{picture}(60,60)(-30,-30)
\thicklines \put(0,0){\line(-1,-1){25}} \put(0,0){\line(1,-1){25}}
\put(0,0){\line(0,1){25}} \put(-25,-27){\makebox(0,0)[t]{Shell 1}}
\put(25,-27){\makebox(0,0)[t]{Shell 2}} \put(0,27){\makebox(0,0)[b]{Shell 3}}
\put(-20,11){\makebox(0,0){Region 1}} \put(0,-18){\makebox(0,0){Region 2}}
\put(20,11){\makebox(0,0){Region 3}}
\end{picture}
\vspace{2mm}
\caption{A schematic spacetime diagram representing a totally inelastic collision.  The
  vertical direction is temporal and time progress upward.}
\label{space-time-diagram}
\end{figure}

To start off consider two spherically symmetric concentric shells.  Each of
these shells is assumed to be infinitely thin and their history (before the
collision) is assumed to be represented by separate timelike
hypersurfaces. These hypersurfaces will also be referred as shells
hereafter. Once the collision occurs it is assumed that the initial
shells merge such that a single infinitesimally thin shell is formed. 
As is depicted by the simple schematic picture in 
figure \ref{space-time-diagram} the spacetime may be divided into three
characteristically different regions. 
In advance to the collision shell 1 and shell 2 indicate the interior and
exterior shells, respectively, while the shell yielded by the merging of these
two shells is represented by shell 3.  The inside spacetime region which has
shells 1 and 3 on its the boundary is called region 1, the region between
shells 1 and 2 is called region 2, while the  outside region bounded by shells
2 and 3 is called region 3.  For notational convenience we introduce region 4
which is identical to region 1.  Hereafter we use capital Latin index $K$ to
indicate the label either of a shell or of a region. Accordingly, the
value of $K$ runs from 1 to 3 unless otherwise stated. 

The metric in the disjoint spherical symmetric regions, labelled by $K$, will
be assumed to be given as 

\begin{equation}\label{met1}
\rmd s_K^2 = - f_K(r)\rmd t^2 + f_K(r)^{-1}\rmd r^2 + r^2\rmd\Omega^2 \, ,
\end{equation}

\noindent 
with

\begin{equation}\label{met2}
f_K(r) = 1 - \frac{2M_K(r)}{r} \; .
\end{equation}

\noindent
Note that this class is generic enough to include both the
Schwarzschild and de Sitter vacuum spacetimes, but the electrovacuum Reisner-Nordstr\"om solution 
does also fit to this form.

In restricting our attention to a system consisting of a gravastar and a
spherical shell falling onto it---the same system investigated in this
paper---in regions 2 and 3, in virtue of Birkhoff's theorem, the  metric has
to be (locally) isometric to that of the  Schwarzschild
spacetime. Accordingly, in regions 2 and 3 the metric can be given by
(\ref{met1}) and (\ref{met2}) with $M_K(r)=M_K$, where the constants $M_K$
denote the corresponding gravitational mass parameters of the pertinent
Schwarzschild spacetimes.  In addition, in region 1 the metric is (locally)
isometric to the de Sitter spacetime with $M_1(r)=(4\pi/3)\rho_0\,r^3$, where
$\rho_0$ is the energy density in the de  Sitter vacuum. In this case,
$M_3$ stands for the total gravitational mass of the system while $M_2$
denotes the total gravitational mass of the entire system.  

Note, however, that in carrying out the following calculations there is no
need to make any specific choice, i.e. the metric in region 1 will only be
required to possess the form (\ref{met1}) and (\ref{met2}). This provides us the 
freedom that whenever
the interior is chosen to be de Sitter type the corresponding equations will immediately 
be applicable to a system formed by a shell falling onto a gravastar, while
if the interior is chosen to be Schwarzschild type the relevant equations
describe the inelastic collision of two confocal spherical shells in a
Schwarzschild vacuum spacetime.

In the following denote by $u_K^a$ the 4-velocity of shell $K$.  The components 
of the 4-velocity, with respect to the coordinates in regions bounding shell $K$, 
can be given as

\begin{equation}
\label{uup}
u_{K(\pm)}^\alpha = \left(\frac{\rmd t_{K(\pm)}}{\rmd \tau_K},\frac{\rmd
  r_K}{\rmd \tau_K},0,0\right),
\end{equation}

\noindent
where $\tau_K$ denotes the proper time on the shell $K$ and the suffixes $(+)$ or
$(-)$ indicate whether the pertinent region is outside or inside with respect
to the shell  $K$.  Note that---as opposed to the area radius, $r$, which is everywhere 
continuous---the time coordinate is not continuous across a shell.
Once the components of $u_K^a$ are fixed the spacelike outward pointing unit
vector, $n_K^a$, normal to the shell $K$ can be determined by the
orthogonality condition, $n_{K}{}_a u_{K}^a = 0$, and its components can
be given as

\begin{equation}
\label{nup}
n_{K(\pm)}^\alpha = \left(\frac{\rmd
  r_K/\rmd\tau_K}{f_{(\pm)}(r)},f_{(\pm)}(r)\frac{\rmd
  t_{K(\pm)}}{\rmd\tau_K},0,0\right).
\end{equation}

The evaluation of (\ref{uup}) and (\ref{nup}) requires the determination 
of the radial velocity, $\rmd r_K/\rmd\tau_K$. There are various ways to get the value of 
$\rmd r_K/\rmd\tau_K$. The most widely known method is based on the
thin-shell formalism of Israel-Sen-Lanczos-Darmois.  For a comprehensive summary 
of this formalism see \cite{Visser04}. After an appropriate reformulation of the `master equation' (see
equation (38) of \cite{Visser04}), which is in fact
an energy balance equation, $\rmd r_K/\rmd\tau_K$, can be given as  

\begin{equation}
\label{rdot}
\left(\frac{\rmd r_K}{\rmd\tau_K}\right)^2 = \frac{g_K(r)^2}{m_K(r)^2} - 1 +
\frac{M_K(r) + M_{K+1}(r)}{r} + \frac{m_K(r)^2}{4r^2} \; , 
\end{equation}

\noindent
where $m_K(r)$ denotes the rest mass, while $g_K(r)$ stands for the
gravitational mass of the shell $K$. Accordingly, $g_K(r) = M_{K+1}(r) - M_K(r)$
for $K=1,2$ and $g_3(r) = M_3(r) - M_1(r)$.  Note that the rest mass of the
shell of a gravastar, in general, is a function of the area radius,
nevertheless, the rest mass of a dust shell is constant.

Recall that the components of the 4-velocity cannot be independent as $u_{K}{}_a u_{K}^a = -1$ which, in particular, implies that 

\begin{equation}
\label{tdot}
\frac{\rmd t_{K(\pm)}}{\rmd\tau_K} = \frac{\sqrt{(\rmd r_K/\rmd\tau_K)^2 +
    f_{(\pm)}(r)}}{f_{(\pm)}(r)} \; .
\end{equation}
 
\noindent
The substitution of (\ref{rdot}) into (\ref{tdot}), and some
straightforward algebraic manipulations, yields that 

\begin{equation}
\label{A7}
\frac{\rmd t_{K(\pm)}}{\rmd\tau_K} = \frac{g_K(r) \mp
  h_K(r)}{m_K(r)f_{K(\pm)}(r)} \; ,
\end{equation}

\noindent
where $h_K(r) = m_K^2(r)/2r$, referred as the self-gravity of the shell $K$.

\bigskip

Returning to the main line of the argument recall that our aim is to determine
the motion of the shell yielded by the merging of the two initial ones. In
doing so the only guiding principle we can use is the conservation of
4-momentum 

\begin{equation}
\label{4-momentum}
m_3u_3^a = m_1u_1^a + m_2u_2^a \; ,
\end{equation}

\noindent
which has to be evaluated at the moment of the collision. Note that the
content of (\ref{4-momentum}) is nothing but two scalar equations. The
temporal component corresponds to the energy conservation, while the radial
component expresses to the radial momentum conservation. The other two
components vanish identically due to the spherical symmetry.  The two
non-trivial equations completely determine the motion of the outgoing shell
since, as we shall see below, both the radial velocity and the rest mass $m_3$
get to be fixed by them. Note that since the collision is inelastic a part of
the kinetic energy is converted into binding energy which, in particular,
implies that $m_3\neq m_1+m_2$.

In applying (\ref{4-momentum}) each of the 4-velocities have to be expressed with respect to a common
coordinate basis field. Start by choosing the coordinate basis field of region 1
as our reference frame. Then, in virtue of (\ref{uup}), the formal expressions of the components of $u_1^a$ and $u_3^a$ can 
immediately be determined. Similarly, in virtue of (\ref{uup}) and (\ref{nup}), the
coordinate basis components of $u_1^a$ and $n_1^a$ of region 1 are known, therefore $u_2^a$ may be
expressed there as the linear combination

\begin{equation}
\label{u2a}
u_2^a = -(u_2^bu_{1b})\,u_1^a + (u_2^bn_{1b})\,n_1^a\,.
\end{equation}

\noindent
By applying (\ref{met1}), (\ref{met2}), (\ref{uup}) and (\ref{nup}) 

\begin{eqnarray}
\label{ndn}
n_{K(\pm)\alpha} & = & \left(-\frac{\rmd r_K}{\rmd\tau_K},\frac{\rmd
  t_{K(\pm)}}{\rmd\tau_K},0,0\right) , \\
\label{udn}
u_{K(\pm)\alpha} & = & \left(-f_{(\pm)}(r)\frac{\rmd
  t_{K(\pm)}}{\rmd\tau_K},\frac{\rmd r_K/\rmd\tau_K}{f_{(\pm)}(r)},0,0\right),
\end{eqnarray}

\noindent
which, along with the notation $p_K=m_K(\rmd r_K/\rmd\tau_K)$, implies that  

\begin{eqnarray}
\label{scalarprduct1}
u_2^bu_{1b} & = & \frac{p_1p_2 - (g_1 - h_1)(g_2 + h_2)}{m_1m_2f_2}\;, \\
\label{scalarprduct12}
u_2^bn_{1b} & = &\frac{(g_1 - h_1)p_2 - (g_2 + h_2)p_1}{m_1m_2f_2}\;.
\end{eqnarray}

\noindent
By substituting these relations into (\ref{u2a}) 
the components of $u_2^a$ in region 1 are determined. Finally, by making
use of these components, along with (\ref{A7}), the two non-trivial equations in
(\ref{4-momentum}) can be given as 

\begin{eqnarray}
\label{w3}
&& h_3 - g_3 =  h_1 - g_1 - \frac{(g_2 + h_2)(p_1^2 - g_1^2 + h_1^2) +
  2p_1h_1p_2}{m_1^2f_2}\; , \\
\label{p3}
&& \phantom{h_3 - \,\,} p_3  =  p_1 - \frac{p_2(p_1^2 - g_1^2 + h_1^2) + 2p_1h_1(g_2 +
  h_2)}{m_1^2f_2}\; ,
\end{eqnarray}

\noindent
where $g_3$ can be eliminated from (\ref{w3}) by using the relation $g_3=g_1 + g_2$.  We would like to
emphasize that in deriving these relations no use of the EOS of the shells has been made. 

Let us finally restrict our considerations to the case of a dust shell falling
onto a gravastar.  Since the gravastar is in equilibrium $p_1=0$.
Thus, in virtue of (\ref{p3}), we get

\begin{equation}
\label{forgravastar}
m_3v_3 = m_2v_2\,\frac{g_1^2 - h_1^2}{m_1^2f_2} \; . \\
\end{equation}

\noindent
Whenever $v_2=0$ then $v_3=0$ also holds. Whenever $v_2\not=0$ and the gravitational mass of the shell is equal to its proper mass---in this case $v_2(\infty)=0$---by making use of the notation of section 3 the relations $m_2 = \varepsilon M_2$, $m_3 = 4\pi(\alpha_02M_2)^2\Sigma_3/M_2$, $f_2 = 1-1/\alpha_0$, $g_1 = M_2(1-\eta\alpha_0^3)$ and (\ref{v2}) can also hold. By substituting them into (\ref{forgravastar}) the radial velocity, $v_3 $, of the shell, at the moment of the collision, can be written as 
 
\begin{equation} 
\label{appendixv3}
v_3 =
-\frac{4\pi\,\Sigma_0^2\,(\alpha_0-1)\,\varepsilon\,\sqrt{\varepsilon^2
    +8\,\varepsilon\,\alpha_0+16\,\alpha_0}} 
{\Sigma_3\,(1-2\,\eta\,\alpha_0^3+\eta^2\alpha_0^6
  -4096\,\pi^4\alpha_0^6\,\Sigma_0^4)}\,. \\ 
\end{equation}

\end{document}